\documentstyle[11pt,amssymb]{article}

\makeatletter
\@addtoreset{equation}{section}
\makeatother

\def\dalemb#1#2{{\vbox{\hrule height .#2pt
        \hbox{\vrule width.#2pt height#1pt \kern#1pt
                \vrule width.#2pt}
        \hrule height.#2pt}}}

\def\cA{{\cal A}}
\def\cT{{\cal T}}
\def\B{{\cal B}}

\def\hi{\hat{\imath}}
\def\hj{\hat{\jmath}}

\def\0{{\sst{(0)}}}
\def\1{{\sst{(1)}}}
\def\2{{\sst{(2)}}}
\def\3{{\sst{(3)}}}
\def\4{{\sst{(4)}}}
\def\5{{\sst{(5)}}}
\def\6{{\sst{(6)}}}
\def\7{{\sst{(7)}}}
\def\8{{\sst{(8)}}}
\def\n{{\sst{(n)}}}

\def\Z{\rlap{\sf Z}\mkern3mu{\sf Z}}

\def\G{{\cal G}}

\def\CC{{\cal C}}
\def\S{{\cal S}}
\def\P{{\cal P}}
\def\cD{{\cal D}}

\def\cO{{\cal O}}
\def\cL{{\cal L}}

\def\cX{{\cal X}}

\let\a=\alpha \let\b=\beta \let\g=\gamma \let\d=\delta \let\e=\epsilon
 \let\h=\eta \let\q=\theta  \let\k=\kappa
\let\l=\lambda \let\m=\mu \let\n=\nu   \let\r=\rho
\let\s=\sigma \let\t=\tau   

\let\w=\omega  \let\D=\Delta  \let\L=\Lambda
      \let\vart=\vartheta
    \let\T=\Theta  \let\G=\Gamma
\let\vare=\varepsilon 
   \let\Si=\Sigma
   
\def\nn{\nonumber} \def\bd{\begin{document}} \def\ed{\end{document}}
\def\ds{\documentstyle} \let\fr=\frac \let\bl=\bigl \let\br=\bigr
\def\HW{Ho\v{r}ava-Witten }
\def\WZW{Wess-Zumino-Witten }
\let\Br=\Bigr \let\Bl=\Bigl 
\let\na=\nabla
\let\pa=\partial \let\ov=\overline 
\newcommand{\be}{\begin{equation}} 
\newcommand{\ee}{\end{equation}} 
\def\ba{\begin{array}}
\def\ea{\end{array}}
\def\ft#1#2{{\textstyle{{\scriptstyle #1}\over {\scriptstyle #2}}}}
\def\fft#1#2{{#1 \over #2}}
\def\del{\partial}
\def\sst#1{{\scriptscriptstyle #1}}
\def\oneone{\rlap 1\mkern4mu{\rm l}}
\def\ie{{\it i.e.\ }}
\def\etc{{\it etc.\ }}
\def\via{{\it via}}
\def\semi{{\ltimes}}
\def\cV{{\cal V}}
\def\str{{\rm str}}
\def\jm{{\rm j}}
\def\im{{\rm i}}
\def\mapright#1{\smash{\mathop{-\!\!\!-\!\!\!-\!\!\!-\!\!\!-\!\!\!
             \longrightarrow}\limits^{#1}}}
\def\maprightt#1#2{\smash{\mathop{-\!\!\!-\!\!\!-\!\!\!-\!\!\!-\!\!\!
             \longrightarrow}\limits^{#1}_{#2}}}
\def\bb#1{{\mbox{\boldmath $#1$}}}

\def\ul#1{{\underline{#1}}}
\def\uh#1{{\underline{\hat{#1}}}}

\def\cramp{\medmuskip = 2mu plus 1mu minus 2mu}
\def\cramper{\medmuskip = 2mu plus 1mu minus 2mu}
\def\crampest{\medmuskip = 1mu plus 1mu minus 1mu}
\def\uncramp{\medmuskip = 4mu plus 2mu minus 4mu}

\newcommand{\ho}[1]{$\, ^{#1}$}
\newcommand{\hoch}[1]{$\, ^{#1}$}
\newcommand{\bea}{\begin{eqnarray}} 
\newcommand{\eea}{\end{eqnarray}} 
\newcommand{\ra}{\rightarrow}
\newcommand{\lra}{\longrightarrow}
\newcommand{\Lra}{\Leftrightarrow}
\newcommand{\ap}{\alpha^\prime}
\newcommand{\bp}{\tilde \beta^\prime}
\newcommand{\tr}{{\rm tr} }
\newcommand{\Tr}{{\rm Tr} } 
\newcommand{\NP}{Nucl. Phys. }
\newcommand{\tamphys}{\it Center for Theoretical Physics\\
Texas A\&M University, College Station, Texas 77843}
\newcommand{\ens}{\it Laboratoire de Physique Th\'eorique de l'\'Ecole
Normale Sup\'erieure\hoch{2,3}\\
24 Rue Lhomond - 75231 Paris CEDEX 05}
\newcommand{\upenn}{\it \hoch{1}David Rittenhouse Laboratory\\
Department of Physics and Astronomy\\
University of Pennsylvania, Philadelphia, Pennsylvania 19104}
\newcommand{\curie}{\it \hoch{3}Institut Henri Poincar\'e\\
Universit\'e Pierre et Marie Curie\\
75231 Paris, CEDEX 05}
\newcommand{\ufc}{\it \hoch{2}Departamento de Fisica\\
Universidade Federal do Cear\'a, Fortaleza, Brazil}

\newcommand{\auth}{Eduardo Lima\hoch{1,2}, Burt Ovrut\hoch{1,3} and
Jaemo Park\hoch{1}}

\def\hk{\hat{k}}
\def\hM{\hat{M}}
\def\hN{\hat{N}}
\def\hO{\hat{O}}
\def\hP{\hat{P}}
\def\hA{\hat{A}}
\def\hB{\hat{B}}
\def\hC{\hat{C}}
\def\hD{\hat{D}}
\def\hE{\hat{E}}
\def\ha{\hat{a}}
\def\hb{\hat{b}}
\def\hc{\hat{c}}
\def\hbY{\hat{\Bbb{Y}}}
\def\hbZ{\hat{\Bbb{Z}}}
\def\hbA{\hat{\Bbb{A}}}
\def\hbC{\hat{\Bbb{C}}}
\def\hbB{\hat{\Bbb{B}}}
\def\hbD{\hat{\Bbb{D}}}
\def\hbM{\hat{\Bbb{M}}}
\def\hbN{\hat{\Bbb{N}}}
\def\hbO{\hat{\Bbb{O}}}
\def\hbP{\hat{\Bbb{P}}}
\def\hbQ{\hat{\Bbb{Q}}}
\def\hbH{\hat{\Bbb{H}}}
\def\hbE{\hat{\Bbb{E}}}
\def\hbF{\hat{\Bbb{F}}}
\def\hbG{\hat{\Bbb{G}}}
\def\uhbY{\underline{\hat{\Bbb{Y}}}}
\def\uhbZ{\underline{\hat{\Bbb{Z}}}}
\def\uhbA{\underline{\hat{\Bbb{A}}}}
\def\uhbC{\underline{\hat{\Bbb{C}}}}
\def\uhbB{\underline{\hat{\Bbb{B}}}}
\def\uhbD{\underline{\hat{\Bbb{D}}}}
\def\uhbM{\underline{\hat{\Bbb{M}}}}
\def\uhbN{\underline{\hat{\Bbb{N}}}}
\def\uhbP{\underline{\hat{\Bbb{P}}}}
\def\uhbQ{\underline{\hat{\Bbb{Q}}}}
\def\uhbH{\underline{\hat{\Bbb{H}}}}
\def\uhbE{\underline{\hat{\Bbb{E}}}}
\def\uhbF{\underline{\hat{\Bbb{F}}}}
\def\uhbG{\underline{\hat{\Bbb{G}}}}
\def\BbM{{\Bbb{M}}}
\def\BbN{{\Bbb{N}}}
\def\BbO{{\Bbb{O}}}
\def\BbP{{\Bbb{P}}}
\def\BbQ{{\Bbb{Q}}}
\def\BbF{{\Bbb{F}}}
\def\BbA{{\Bbb{A}}}
\def\BbL{{\Bbb{L}}}
\def\BbZ{{\Bbb{Z}}}
\def\BbY{{\Bbb{Y}}}
\def\BbB{{\Bbb{B}}}
\def\BbC{{\Bbb{C}}}
\def\BbD{{\Bbb{D}}}
\def\BbH{{\Bbb{H}}}
\def\BbV{{\Bbb{V}}}
\def\BbE{{\Bbb{E}}}
\def\ubM{{\underline{\Bbb{M}}}}
\def\ubN{{\underline{\Bbb{N}}}}
\def\ubP{{\underline{\Bbb{P}}}}
\def\ubQ{{\underline{\Bbb{Q}}}}
\def\ubR{{\underline{\Bbb{R}}}}
\def\ubS{{\underline{\Bbb{S}}}}
\def\ubT{{\underline{\Bbb{T}}}}
\def\ubF{{\underline{\Bbb{F}}}}
\def\ubA{{\underline{\Bbb{A}}}}
\def\ubL{{\underline{\Bbb{L}}}}
\def\ubY{{\underline{\Bbb{Y}}}}
\def\ubZ{{\underline{\Bbb{Z}}}}
\def\ubB{{\underline{\Bbb{B}}}}
\def\ubC{{\underline{\Bbb{C}}}}
\def\ubD{{\underline{\Bbb{D}}}}
\def\ubH{{\underline{\Bbb{H}}}}
\def\ubV{{\underline{\Bbb{V}}}}
\def\ubE{{\underline{\Bbb{E}}}}
\def\bm{\bar{m}}
\def\bn{\bar{n}}
\def\bz{\bar{z}}
\def\delslash{\del\!\!\!/}
\def\Dslash{D\!\!\!\!/}
\def\Oslash{\cO\!\!\!\!/}

\begin{document}
\begin{flushright}
\hfill{UPR 926T}\\
\hfill{hep-th/0102046}\\
\hfill{February 2001}\\
\end{flushright}


\begin{center}
{ \large {\bf Five-Brane Superpotentials
in Heterotic $M$-theory  }}

\vspace{15pt}
\auth

\vspace{15pt}

{\upenn}

\vspace{5pt}

{\ufc}

\vspace{5pt}

{\curie}

\vspace{10pt}

\vspace{40pt}

\underline{ABSTRACT}
\end{center}

The open supermembrane contribution to the non-perturbative superpotential
of bulk space five-branes in heterotic $M$-theory is presented. We explicitly 
compute the superpotential for the modulus associated with the separation
of a bulk five-brane from an end-of-the-world three-brane. The gauge 
and $\kappa$-invariant boundary strings of such open supermembranes are 
given and the role of the holomorphic vector bundle on the orbifold fixed 
plane boundary is discussed in detail.

{\vfill\leftline{}\vfill
\footnoterule

{\footnotesize \hoch{ } Research supported in part by DOE grant
DE-AC02-76ER03071 \vskip -12pt} \vskip 14pt

\pagebreak
\setcounter{page}{1}

\section{Introduction}

One of the principal problems obstructing attempts to obtain the standard
model from $M$-theory, that is, how to obtain chiral fermions in the low energy
effective theory, was solved in several papers by \HW \cite{HorWit1,HorWit2} and
Witten \cite{W97}. These authors compactified eleven-dimensional $N=1$ supergravity on
the orbifold $S^1/\Z_2$, producing chiral fermions on each of the two
ten-dimensional orbifold fixed planes. They then showed that cancellation of
the gravitational anomalies induced by these chiral fermions uniquely requires
that each orbifold fixed plane supports an $N=1$, $E_{8}$ super-Yang-Mills
multiplet. Ho\v rava-Witten theory, therefore, is a theory with an
eleven-dimensional ``bulk'' space bounded on two sides by ten-dimensional
$S^1/\Z_2$ orbifold planes. The relation of this theory to $N=1$
four-dimensional theories was explored in \cite{W97,c10,BAD}, but these papers
compactified Ho\v rava-Witten theory directly to four-dimensions.

In a series of papers \cite{B34}--\cite{B31}, it was shown that it is natural 
to compactify \HW theory, not directly to four-dimensions, but, rather, 
on a Calabi-Yau threefold to a theory with a five-dimensional
bulk space bounded, on each end of the orbifold, by four-dimensional
BPS three-branes. In this compactification, called heterotic $M$-theory, our
observable world arises as the worldvolume theory of one of the boundary
three-branes, the other boundary brane forming a hidden sector. Heterotic
$M$-theory represents a fundamental realization of a ``brane universe'' directly
from $M$-theory. It was demonstrated explicitly in \cite{B27}--\cite{B3} 
that both grand unified
theories of particle physics and the standard model can arise on the
observable three-brane by appropriately specifying semi-stable, holomorphic
$E_{8}$ vector bundles on the associated Calabi-Yau space. These Yang-Mills
``instantons'' break $E_{8}$ to phenomenologically interesting gauge groups,
such as $SU(3)\times SU(2)\times U(1)$, and lead to three families of quarks
and leptons. However, it was shown in these papers that, generically, anomaly
cancellation requires the existence of BPS five-branes, wrapped on holomorphic
curves in the Calabi-Yau threefold, in the bulk space. These five-branes
represent new, non-perturbative physics that might have dramatic effects both
in low energy particle physics and in 
cosmology \cite{c32}--\cite{c45}. 
Therefore, it is of
importance to have a detailed understanding of their dynamics.

With this in mind, in a recent paper \cite{LOPR}, we computed the superpotential 
induced at low energy by the exchange of open supermembranes between the 
two orbifold fixed planes. This superpotential is an explicit holomorphic 
function of the $(1,1)$-moduli of the Calabi-Yau threefold. In addition, we
showed that this superpotential is only non-vanishing under restrictive
topological conditions on the end-of-the-world orbifold plane instantons, 
namely, that the restriction of each vector bundle to the holomorphic 
curve around which the supermembrane is wrapped be trivial. In this paper, we
extend these results to compute the low energy $N=1$ superpotential induced by
the exchange of open supermembranes between one end-of-the-world BPS 
three-brane and a wrapped five-brane in the bulk space. This calculation,
although related to the one performed for the two orbifold fixed planes, has
many features that are unique to the bulk space five-brane. We find that the
superpotential is a holomorphic function of a new, composite modulus. This
modulus is a specific combination of the translation modulus of the five-brane, the
real and imaginary parts of the $(1,1)$-modulus associated with the
holomorphic curve on which the five-brane is wrapped and the ``axion''
modulus, which is related to the worldvolume two-form of the five-brane. Again,
we find that this superpotential is only non-vanishing if 
the vector bundle associated with the end-of-the-world
three-brane is trivial when restricted to the holomorphic curve on which the
five-brane and open supermembrane are wrapped.

Specifically, we do the following. In Section 2 we discuss supermembranes and
five-branes in both eleven-dimensional supergravity and in Ho\v rava-Witten
theory. The $\kappa$-invariant action for an open supermembrane with one boundary
string on an orbifold fixed plane and the other on a bulk space five-brane is
studied in detail in Section 3. Section 4 is devoted to a discussion of the
compactification of this theory on a Calabi-Yau threefold to heterotic
$M$-theory and the further dimensional reduction on the $S^1/\Z_2$ orbifold.
The effective action is shown to reduce to that of the heterotic superstring
coupled to one $E_{8}$ gauge background, a Neveu-Schwarz five-brane and wrapped
on a holomrphic curve in the Calabi-Yau manifold. In Section 5, we review the
relevant moduli in heterotic $M$-theory and their reduction to the
four-dimensional effective theory. We discuss in detail the $(1,1)$-modulus
$\cT$ associated with the holomorphic curve and the translational 
chiral multiplet $\bf Y \rm$ that will appear in the superpotential. We also
discuss the method for calculating the superpotential from the open
supermembrane contribution to the relevant fermion two-point function. Section
6 is devoted to the explicit calculation of the non-perturbative corrections
to this two-point function using a saddle point approximation. We present a
careful discussion of gauge fixing and zero-modes, calculate the bosonic
holomorphic contribution and compute the formal expressions for the determinants
associated with quadratic fluctuation terms. In Section 7 we calculate the
Wess-Zumino-Witten determinant related to quadratic fluctuations in
the background of an $E_{8}$ gauge instanton. It is shown that this
determinant is only non-vanishing if the restriction of the holomorphic vector
bundle to the curve on which the heterotic string is wrapped is trivial.
Finally, we extract the complete expression for the superpotential associated
with the five-brane translation modulus in Section 8. Our notation and
conventions are discussed in the Appendix.

Our work, both in this paper and in \cite{LOPR}, is based on the ground-breaking
formalism presented in \cite{BeckerBS,HarvMoor}. Recently, a paper due to 
Moore, Peradze and Saulina \cite{MPS} appeared which studied topics similar to 
those presented here and in \cite{LOPR}.
Some of our results are similar to theirs and much is new or complementary. We
acknowledge their work and appreciate their pre-announcement of our
independent study of this subject. We want to point out and emphasize the
paper of Derendinger and Sauser \cite{Derend} on the perturbative low energy 
effective 
theory of five-branes in heterotic $M$-theory. These authors elucidated the 
relevant moduli associated with five-brane dynamics and computed their 
contribution to the four-dimensional K\"{a}hler potential. In this paper, we add  
the non-perturbative superpotential contributions. We note that the same
moduli naturally arise in our calculation, in a very different context.

\section{Membranes and Five-Branes in \HW Theory:}

\subsection*{Eleven-Dimensional Supergravity, Membranes and Five-Branes:}

$N=1$ supersymmetry in eleven-dimensions has 32 supercharges and 
consists of a single
supergravity multiplet \cite{CJS} containing as its component fields a
graviton $\hat{g}_{\hat{M}\hat{N}}$, a three-form $\hat{C}_{\hat{M}\hat{N}
\hat{P}}$ and a Majorana gravitino $\hat{\Psi}_{\hat{M}}$. The field
strength
of the three-form, defined by $\hat{G}=d\hat{C}$, has as its components
$\hat{G}_{\hat{M}\hat{N}\hat{P}\hat{Q}}=
24\del_{[\hat{M}}\hat{C}_{\hat{N}\hat{P}\hat{Q}]}$. We denote the
coordinates of the real eleven manifold $M_{11}$ as
$(\hat{x}^{\hat{0}},\ldots,\hat{x}^{\hat{9}},\hat{x}^{\hat{11}})$.
The associated action is invariant under the
supersymmetry transformations of the component fields. For our purposes,
we need only specify the supersymmetry variation of the gravitino field 
$\hat{\Psi}_{\hat{M}}$, which is given by
\be
\d_{\hat{\vare}} 
\hat{\Psi}_{\hat{M}} = \hat{D}_{\hat{M}} \hat{\vare} +
\frac{\sqrt{2}}{288}
(\hat{\G}_{\hat{M}}^{\; \hat{N}\hat{P}\hat{Q}\hat{R}} - 8 
\d_{\hat{M}}^{\hat{N}} \hat{\G}^{\hat{P}\hat{Q}\hat{R}} ) \hat{\vare}
\hat{G}_{\hat{N}\hat{P}\hat{Q}\hat{R}} + \cdots , \label{11dSusy}
\ee
where $\hat{\vare}$ is the Majorana supersymmetry parameter and the dots
denote terms that involve the fermion fields of the theory.
The eleven-dimensional spacetime Dirac matrices $\hat{\G}_{\hat{M}}$ satisfy
$\{\hat{\G}_{\hat{M}},\hat{\G}_{\hat{N}}\}=2\hat{g}_{\hat{M}\hat{N}}$.
$N=1$
eleven-dimensional supergravity can be formulated in a superspace with
coordinates
\begin{equation}
\hat{z}^{\BbM}=(\hat{x}^{\hat{M}},\hat{\theta}^{\hat{\mu}}),
\label{eq:burt1}
\end{equation}
where $\hat{x}^{\hat{M}}$, $\hat{M}=0,\ldots,9,11$ are the bosonic
coordinates
introduced above and $\hat{\theta}^{\hat{\mu}}$, $\hat{\mu}=1,\ldots,32$ 
are anti-commuting coordinates in a thirty-two component Majorana spinor.
In this formulation, the graviton and three-form appear as the lowest
components of the super-elfbein $\hbE_{\hbM}^{\;\hbA}$ and the
super-three-form
$\hbC_{\hbC \hbB \hbA}$ respectively. The gravitino arises at order
$\hat{\theta}$ in the expansion of $\hbE_{\hbM}^{\;\hbA}$.

It is well-known that there is a $2+1$-dimensional ``electrically
charged'' membrane solution of the $M$-theory equations of motion that 
preserves one-half of the supersymmetries \cite{DuffStelle}, that is,
16 supercharges. 
The worldvolume action for this supermembrane coupled to background
eleven-dimensional supergravity is known \cite{BST}. It is given, in the
target superspace formulation, by
\be
S_{S\!M} =- T_M \int_{\Si} d^3 \hat{\s} ( 
\sqrt{-\det \hat{g}_{\hi\hj}} - \frac{1}{6} 
\hat{\vare}^{\hi\hj \hat{k}} \hat{\Pi}_{\hi}^{\, \hbA}
\hat{\Pi}_{\hj}^{\, \hbB} \hat{\Pi}_{\hat{k}}^{\, \hbC} 
\hbC_{\hbC\hbB\hbA} ), \label{SMaction}
\ee
where 
\be
T_M = (2 \pi^2 / \k^2)^{1/3} \label{2btension}
\ee
is the membrane tension of mass dimension three and
$\hat{\sigma}^{\hi},\hi =0,1,2$ are the intrinsic
coordinates of the membrane worldvolume $\Si$. Parameter $\k$ is
the eleven-dimensional Newton constant. Furthermore,
\be
\hat{g}_{\hi\hj} = \hat{\Pi}_{\hi}^{\, \hat{A}} \hat{\Pi}_{\hj }^{\,
\hat{B}}
\h_{\hat{A}\hat{B}}, \ \ \ \ \ \ \ 
\hat{\Pi}_{\hi}^{\, \hbA} = \del_{\hi}\hbZ^{\hbM} 
\hbE_{\hbM}^{\; \hbA}, \label{supermetric}
\ee
where $\hbZ^{\hbM}$ represents the superembedding $\hbZ : \Si^{3|0} \ra
M^{11|32}$, 
whose bosonic and fermionic component fields are the background
coordinates
\be
\hbZ^{\hbM}(\hat{\s}) = (\hat{X}^{\hat{M}}(\hat{\s}) , 
\hat{\T}^{\hat{\m}} (\hat{\s}) ), \label{supercoord}
\ee
respectively.
The action is a sigma-model since the super-elfbeins 
$\hbE_{\hbM}^{\; \hbA}$ and the super-three-form
$\hbC_{\hbC\hbB\hbA}$ both depend on the superfields 
$\hbZ^{\hbM}$. The super-elfbeins have, as their first bosonic and
fermionic 
component in the $\hat{\T}$ expansion, the bosonic elfbeins 
$\hat{E}_{\hat{M}}^{\, \hat{A}}$
and the gravitino $\hat{\Psi}_{\hat{M}}^{\; \hat{\a}}$
respectively, while the super-three-form has the bosonic three-form from 
eleven-dimensional supergravity as its leading field component.

The fact that the membrane solution of
$M$-theory preserves one-half of the supersymmetries
translates, when speaking in supermembrane worldvolume language, into the
fact
that the action (\ref{SMaction}) exhibits a local fermionic invariance,
$\kappa$-invariance, that is used to gauge away half of the fermionic 
degrees of freedom. Specifically, the supermembrane action is invariant 
under the local fermionic symmetries
\be
\d_{\hat{\k}} \hat{\T} = 2 \hat{P}_+ \hat{\k} + \cdots, \ \ \ \ \ 
\d_{\hat{\k}} \hat{X}^{\hat{M}} = 2 \bar{\hat{\T}} 
\hat{\G}^{\hat{M}} \hat{P}_+ \hat{\k} + \cdots , \label{kappa}
\ee
where $\hat{\k} (\hat{\s})$ is an eleven-dimensional local spinor
parameter and $\hat{P}_{\pm}$ are the projection operators
\be
\hat{P}_{\pm} \equiv \frac{1}{2} (1 \pm \frac{1}{6\sqrt{-\det
\hat{g}_{\hi\hj}}} 
\hat{\vare}^{\hi\hj \hat{k}} \hat{\Pi}_{\hi}^{\, \hat{A}} 
\hat{\Pi}_{\hj }^{\, \hat{B}} \hat{\Pi}_{\hat{k}}^{\, \hat{C}} 
\hat{\G}_{\hat{A}\hat{B}\hat{C}} ), \label{projop}
\ee
obeying
\be
\hat{P}_{\pm}^2 = \hat{P}_{\pm} , \ \ \ \ \hat{P}_+ \hat{P}_- = 0 ,
\ \ \ \ \hat{P}_+ + \hat{P}_- = 1.
\ee
It follows from the first equation in (\ref{kappa}) that the $\hat{P}_+\hat{\T}$
component of spinor $\hat{\T}$ can be transformed away by a $\k$-transformation.
Note that (\ref{kappa}) includes only the leading order terms in
$\hat{\T}$, which is all that is required to discuss the supersymmetry 
properties of the membrane. It can be shown that the membrane action
(\ref{SMaction}) will be
invariant under transformations (\ref{kappa}) if and only if the
background
superfields $\hbE_{\hbM}^{\hbA}$ and $\hbC_{\hbC \hbB \hbA}$ satisfy the
eleven-dimensional supergravity constraint equations. However, the general 
bosonic membrane configuration $\hat{X}(\hat{\s})$ is not invariant under
global supersymmetry transformations
\be
\d_{\hat{\vare}} \hat{\T} = \hat{\vare}, \ \ \ \ \ 
\d_{\hat{\vare}} \hat{X}^{\hat{M}} = \bar{\hat{\vare}} 
\hat{\G}^{\hat{M}} \hat{\T},        \label{glfermTransf}
\ee
where $\hat{\vare}$ is an eleven-dimensional spinor independent of
$\hat{\s}$. Nevertheless, one-half of the supersymmetries will remain
unbroken 
if and only if (\ref{glfermTransf}) can be compensated for 
by a $\k$-transformation with a  suitable parameter $\hat{\k}(\hat{\s})$.
That is
\bea
\d \hat{\T} &=& \d_{\hat{\vare}} \hat{\T} + 
\d_{\hat{\k}} \hat{\T} \nn \\
&=& \hat{\vare} + 2 \hat{P}_+ \hat{\k} (\hat{\s}) = 0. \label{kappaTheta}
\eea
In order for this to be satisfied, a necessary condition is that
\be
\hat{P}_- \hat{\vare} = 0. \label{11dBPS} 
\ee

In addition to the supermembrane, it is well-known that there is a
six-dimensional ``solitonic'' five-brane solution of $M$-theory that
preserves one-half of the supersymmetries \cite{}, that is,
16 supercharges. The worldvolume action
for this five-brane coupled to background eleven-dimensional supergravity
is known \cite{BLNPST}, but it is not necessary to give its explicit form in this
paper. Here, it suffices to note the following. The scale of the action is 
set by $T_5$, which is the five-brane tension of mass dimension six given by
\be
T_5 = (4 \pi / \k^4)^{1/3}.
\ee
It follows from (\ref{2btension}) that the relation between $T_5$ and $T_M$ is
\be
T_5 = \frac{2\pi}{\k^2} \frac{1}{T_M} .
\ee
The five-brane worldvolume $\ul{M}_6$
has the six intrinsic coordinates $\xi^{\ul{r}}, \ul{r}=0,\ldots,5$ and 
worldvolume metric
\be
\ul{g}_{\ul{r}\ul{s}} = \uh{\Pi}_{\ul{r}}^{\,\hat{A}} 
\uh{\Pi}_{\ul{s}}^{\,\hat{B}} \h_{\hat{A}\hat{B}}, \ \ \ \ \ \ \ \ \ \ \ \ \ 
\uh{\Pi}_{\ul{r}}^{\,\hbA} = \del_{\ul{r}}\uhbY^{\hbM} \hbE_{\hbM}^{\;\hbA},
\ee
where $\uhbY^{\hbM}$ represents the superembedding $\uhbY: \ul{M}_6^{6|0} \to 
\hat{M}^{11|32}$ with
\be
\uhbY^{\hbM}(\xi)=(\uh{Y}^{\hat{M}}(\xi),\uh{\Xi}^{\hat{\m}}(\xi)).
\label{5bSE}
\ee
In addition to $\uh{Y}^{\hat{M}}$ and $\uh{\Xi}^{\hat{\m}}$, 
the five-brane theory also requires the introduction of
a worldvolume
two-form, $\ul{D}_{\ul{r}\ul{s}}(\xi)$, whose field strength is anti-self-dual.
Finally, we note that the five-brane action contains explicit couplings to
the super-elfbeins $\hbE_{\hbM}^{\;\hbA}$ and super-three-form 
$\hbC_{\hbC\hbB\hbA}$ of the eleven-dimensional background supergravity.

As for the supermembrane, the five-brane worldvolume action exhibits a
$\k$-invariance that can be used to gauge away half of the fermionic degrees
of freedom. Specifically, the action is invariant under
\be
\d_{\uh{\k}} \uh{\Xi} = 2 \P_+ \uh{\k} + \cdots , \ \ \ \ \ \ \ \ \ \ \ \ \
\d_{\uh{\k}} \uh{Y}^{\hat{M}} = 2 \bar{\uh{\Xi}} \hat{\G}^{\hat{M}} \P_+
\uh{\k} + \cdots , \ \ \ \ \ \ \ \ \ \ \ \ \
\d_{\uh{\k}} \ul{D}_{\ul{r}\ul{s}} = 2\ubC_{\ul{r}\ul{s}\hat{\m}}\P_+
\uh{\k}^{\hat{\m}} ,
\label{5bCoorduk}
\ee
where $\uh{\k} (\xi)$ is an eleven-dimensional local spinor parameter,
\be
\ubC_{\ul{r}\ul{s}\hat{\m}} = \uh{\Pi}_{\ul{r}}^{\,\hbA} 
\uh{\Pi}_{\ul{s}}^{\,\hbB} \hbC_{\hat{\m}\hbB\hbA}
\ee
and $\P_{\pm}$ are projection operators. 
In general, these operators depend in a complicated way on the
three-form
\be
\ubH_{\ul{r}\ul{s}\ul{t}} = (d \ul{D})_{\ul{r}\ul{s}\ul{t}} - 
\ubC_{\ul{r}\ul{s}\ul{t}},
\ee
where
\be
\ubC_{\ul{r}\ul{s}\ul{t}} = \uh{\Pi}_{\ul{r}}^{\,\hbA} 
\uh{\Pi}_{\ul{s}}^{\,\hbB} \uh{\Pi}_{\ul{t}}^{\,\hbC} \hbC_{\hbC\hbB\hbA} 
\ee
is the pullback of the supergravity super-three-form onto the 
five-brane worldvolume. If, however, one chooses
\be
\ubH_{\ul{r}\ul{s}\ul{t}}=0, \label{2.21}
\ee
which we will do for the remainder of this paper,
then these projection operators simplify and are given by
\be
\P_{\pm}= \frac{1}{2} \left(1\pm\frac{1}{6!\sqrt{-\det 
\ul{g}_{\ul{r}\ul{s}}}} \ul{\vare}^{\ul{r}_{1}\ldots\ul{r}_{6}}
\uh{\Pi}_{\ul{r}_{1}}^{\;\hat{A}_{1}}\ldots\uh{\Pi}_{\ul{r}_{6}}^{\;\hat{A}_{6}}
\hat{\G}_{\hat{A}_{1}\ldots\hat{A}_{6}}\right) .
\label{5bPpm}
\ee
Note that (\ref{5bCoorduk}) includes only the leading order terms in 
$\uh{\Xi}$, which is all that is required to discuss the supersymmetry
properties of the five-brane. It can be shown that the five-brane action 
will be invariant under $\k$-transformation (\ref{5bCoorduk}) if and only if
the background superfields $\hbE_{\hbM}^{\;\hbA}$ and $\hbC_{\hbC\hbB\hbA}$
satisfy the constraint equations of eleven-dimensional supergravity.
Using a $\k$-transformation with a suitable parameter $\uh{\k}(\xi)$, the global
supersymmetry transformations
\be
\d_{\uh{\vare}} \uh{\Xi} = \uh{\vare} , \ \ \ \ \ \ \ \ \ \ \ \ \
\d_{\uh{\vare}} \uh{Y}^{\hat{M}} = \bar{\uh{\vare}} \hat{\G}^{\hat{M}}
\uh{\Xi}, \label{5bCoorduvare}
\ee
where $\uh{\vare}$ is an eleven-dimensional spinor independent of $\xi$, can
be compensated for to leave one-half of the supersymmetries unbroken. That is
\be
\d\uh{\Xi}=\uh{\vare}+2\P_{+}\uh{\k}(\xi)=0.
\label{dThetaiszero}
\ee
For this to be satisfied, a necessary condition is that
\be
\P_{-}\uh{\vare}=0. \label{Pmepsiszero}
\ee
It is well-known that after fixing this $\k$-gauge, the 16 unbroken supercharges
arrange themselves as a six-dimensional (2,0) supersymmetry and that
the worldvolume theory
of the five-brane consists of a single tensor multiplet. 
This supermultiplet contains as its component fields
\be
(\ul{D}_{\ul{r}\ul{s}} , \uh{Y}^{\hat{p}} , \chi), \ \ \ \ \ \ \ \ \ 
\ \ \ \ \ \hat{p}=6,\ldots,9,11 , \label{5bsupmult}
\ee
where the field strength of $\ul{D}_{\ul{r}\ul{s}}$ is anti-self-dual, the
five scalars $\uh{Y}^{\hat{p}}$ label the transverse translational modes 
of the five-brane and $\chi$ are the associated fermions.

We now turn to a discussion of supermembranes and five-branes in \HW theory.

\subsection*{Five-Branes and Membranes in \HW Theory:}

When $M$-theory is compactified on $S^1/\Z_2$, it describes the low energy
limit
of the strongly coupled heterotic string theory \cite{HorWit1,HorWit2}. 
We choose $\hat{x}^{\hat{11}}$ as the orbifold direction and parametrize
$S^1$
by $\hat{x}^{\hat{11}} \in [-\pi\r ,\pi\r]$ with the endpoints identified.
The $\Z_2$ symmetry acts by 
further identifying any point $\hat{x}^{\hat{11}}$ with 
$-\hat{x}^{\hat{11}}$ and, therefore, gives rise to two ten-dimensional 
fixed hyperplanes at $\hat{x}^{\hat{11}}=0$ and $\hat{x}^{\hat{11}}=\pi
\r$. We will denote the manifold of either of them by $M_{10}$.
Since, at each 
$\Z_2$ hyperplane, only the field components that are even under the
$\Z_2$
action can survive, the eleven-dimensional supergravity in the bulk 
space is projected
into $N=1$ ten-dimensional chiral supergravity on each boundary. $N=1$
supersymmetry in ten-dimensions preserves 16 supercharges. Furthermore,
cancellation of the chiral anomaly in this theory requires the existence
of 
an $N=1$,
$E_{8}$ super-Yang-Mills multiplet on each fixed hyperplane 
\cite{HorWit1,HorWit2}.
Therefore, the effective action
for $M$-theory on $S^1 / \Z_2$ describes the coupling of two ten-dimensional
$E_8$ super-Yang-Mills theories, one on each hyperplane, to
eleven-dimensional supergravity in the bulk space. 
In order to cancel all chiral anomalies on the hyperplanes, the action 
has to be supplemented by the modified Bianchi 
identity\footnote{The normalization of $\hat{G}$ adopted here differs from 
\cite{HorWit1} by a factor of $\sqrt{2}$ but it agrees with \cite{Ceder},
in which one considers, as we will do in this paper, the superfield
version of the Bianchi identities.}
\be
(d \hat{G})_{\hat{11}MNPQ} = - \frac{1}{4 \pi} 
(\frac{\k}{4 \pi})^{2/3} \left( J^{(1)} \d (\hat{x}^{\hat{11}}) +
J^{(2)} \d (\hat{x}^{\hat{11}} - \pi \r) \right)_{MNPQ} , \label{modBianchi}
\ee
where
\be
J^{(n)} = \mbox{tr} F^{(n)} \wedge F^{(n)} - \frac{1}{2} \mbox{tr} R
\wedge R,
\ee
for $n=1,2$ and $M,N=0,\ldots,9$. 
The solutions to the equations of motion must respect the $\Z_2$ orbifold 
symmetry. For the purposes of this paper, we need only specify the
transformation property of the gravitino under $\Z_2$. This is given by
\be
\hat{\Psi}_{M} (\hat{x}^{\hat{11}}) = \hat{\G}_{\hat{11}}
\hat{\Psi}_{M} (-\hat{x}^{\hat{11}}), \ \ \ \ \ \ \ 
\hat{\Psi}_{\hat{11}} (\hat{x}^{\hat{11}}) = -
\hat{\G}_{\hat{11}} \hat{\Psi}_{\hat{11}} (-\hat{x}^{\hat{11}}).
\label{psiuz2}
\ee
where $\hat{\G}_{\hat{11}}=\hat{\G}_{\hat{0}}\hat{\G}_{\hat{1}}
\cdots\hat{\G}_{\hat{9}}$. In order for the supersymmetry transformations 
of the gravitino to be consistent with the $\Z_2$ symmetry, the 
eleven-dimensional Majorana spinor $\hat{\vare}$ in (\ref{11dSusy}) must
satisfy
\be
\hat{\vare} (\hat{x}^{\hat{11}}) = \hat{\G}_{\hat{11}}
\hat{\vare} (-\hat{x}^{\hat{11}}). \label{epsuz2}
\ee
This equation does not restrict the number of independent components of
the
spinor fields $\hat{\vare}$ at any point in the bulk space. However, at 
each of the $\Z_2$ hyperplanes, constraint (\ref{epsuz2}) becomes the 
ten-dimensional chirality condition
\be
\frac{1}{2} (1 - \hat{\G}_{\hat{11}}) \hat{\vare} = 0, \ \ \ \ \ \  
\mbox{at} \ \ \ \hat{x}^{\hat{11}} = 0, \pi \r. \label{z2chiral}
\ee
This condition implies that the theory exhibits $N=1$ supersymmetry on each
of the ten-dimensional orbifold fixed planes.

Five-brane solutions were explicitly constructed for \HW theory
in \cite{LalLukOvr}. There, the five-brane solution of eleven-dimensional supergravity
was shown to satisfy the equations of motion of the theory subject to the
$\Z_2$ constraints. There are two different ways to orient the five-brane
with respect to the orbifold direction, that is, $\hat{x}^{11}$ can be
either a transverse coordinate or a coordinate oriented in the direction of the
five-brane. In the first case, the five-brane is parallel to the
hyperplanes. In the second case, it intersects each of them along a 4+1 
dimensional brane. Note, however, that there is no BPS 4-brane in 
ten-dimensional heterotic string
theory. It follows that the second orientation cannot preserve supercharges
compatible with the $N=1$
supersymmetry on the boundary hyperplanes. Hence, such five-branes are
not of interest from the point of view of this paper, and we consider only
five-branes parallel to the orbifold fixed planes. We now show that these
parallel five-branes do conserve supercharges compatible with the
boundary hyperplane supersymmetry. As shown above, for any orientation of 
the five-brane, one half of the supersymmetries will remain unbroken if
and only if the target supersymmetry transformation with spinor parameter
$\uh{\vare}$ can be compensated for by a $\k$-transformation with
a suitable parameter $\uh{\k}(\xi)$. For this to be the case $\uh{\vare}$
must satisfy $\P_- \uh{\vare} = 0$, where $\P_-$ is given in (\ref{5bPpm}). 
Now choose the orientation of the fivebrane to be parallel to the orbifold
fixed planes. In this case, we can take the fields $\uh{Y}$ and
$\uh{\Xi}$ such that
\bea
\uh{Y}^{\hat{p}'}&=&\d_{\ul{r}}^{\hat{p}'}\xi^{\ul{r}}, \qquad 
\hat{p}',\ul{r}=0,\ldots,5, \nn \\
\uh{Y}^{\hat{p}}&=&0, \qquad \ \ \ \ \hat{p}=6,\ldots,9, \nn \\
\uh{Y}^{\hat{11}}&=& {Y}, \ \ \ \ \ \ \ \ \ \ \ \ \ \ \ \ 
\ \uh{\Xi} = 0 , \label{5bconfig}
\eea
where ${Y}$ is the location of the five-brane along the orbifold 
direction. Expression (\ref{Pmepsiszero}) then becomes
\be
\P_{-}\uh{\vare}=\frac{1}{2}(1-\hat{\G}_{012345})\uh{\vare}=0 ,
\label{5bchirlty}
\ee
where $\hat{\G}_{012345}=\hat{\G}_{\hat{0}}\cdots\hat{\G}_{\hat{5}}$.
Note that the 16 supercharges preserved on the five-brane worldvolume by
chirality condition (\ref{5bchirlty}) form a
(2,0)-supersymmetry on the five-brane. 

As discussed previously, the
five-brane worldvolume fields form a tensor supermultiplet of
(2,0)-supersymmetry. This contains, among other things, a two-form
$\ul{D}_{\ul{r}\ul{s}}$ whose field strength is anti-self-dual.
It is important to note that the presence of an anti-self-dual tensor in
six-dimensions leads to a gravitational anomaly on the five-brane
worldvolume.
This must be canceled by adding the appropriate higher dimensional
interactions to the eleven-dimensional supergravity theory and by
modifying Bianchi identity (\ref{modBianchi}) to
\be
(d\hat{G})_{\hat{11}MNPQ}=-\frac{1}{4\pi}(\frac{\kappa}{4\pi})^{2/3}
\left(J^{(1)}\delta(\hat{x}^{\hat{11}})+J^{(2)}\delta(\hat{x}^{\hat{11}}-
\pi\rho)+J_5 \delta(\hat{x}^{\hat{11}}-{Y})\right)_{MNPQ},
\label{eq:burt12}
\ee
where $J_5$ is the four-form source that is Poincare dual to the homology
class of the complex curve on which the five-brane is wrapped. 

Now consider supermembranes in the \HW context. We begin by assuming
there are no five-branes in the bulk space.
There are two different ways to orient the membrane with
respect to the orbifold direction, that is,  $\hat{x}^{\hat{11}}$ can
either be a transverse coordinate or a coordinate oriented in the
direction of
the membrane. In the first case, the membrane 
is parallel to the hyperplanes. In the second case, it
extends between the two hyperplanes and intersects each of
them along a $1+1$-dimensional string. This latter configuration is 
sometimes referred to as an open supermembrane. It was shown in \cite{LalLukOvr} that
the parallel configuration cannot preserve supercharges compatible with
the $N=1$ supersymetry on the boundary
hyperplanes. This is readily understood once one notices that a such a
parallel membrane would correspond to a BPS membrane in ten-dimensional
heterotic string theory. However, no such membrane exists and, therefore,
parallel membranes are not of interest in this paper. Henceforth, we only
consider
open supermembrane configurations. We now show that these configurations
do preserve supercharges compatible with the $N=1$ supersymmetry on 
the orbifold planes.

We have seen in (\ref{11dBPS}) that, in order for supersymmetry to be
preserved, the global supersymmetry parameter $\hat{\vare}$ of the
membrane worldvolume theory must satisfy
$\hat{P}_- \hat{\vare}=0$, where $\hat{P}_-$ is given in (\ref{projop}).
An open submembrane  is oriented perpendicular to the ten-dimensional 
hyperplanes. Therefore, we can choose the fields
such that
\bea
\hat{X}^{\hat{0}} &=& \hat{\s}^{\hat{0}}, \ \ \ \ \ \ \ 
\hat{X}^{\hat{1}} = \hat{\s}^{\hat{1}}, \ \ \ \ \ \ \ 
\hat{X}^{\hat{11}} = \hat{\s}^{\hat{2}}, \nn \\
\hat{X}^{\hat{m}} &=& 0, \ \ \ \hat{m}=2,3,\ldots,9 \ \ \ \ \ \ \ 
\hat{\T} = 0, \label{rightconf}
\eea
so that $\hat{P}_- \hat{\vare} = 0$ now becomes
\be
\hat{P}_- \hat{\vare} = \frac{1}{2} (1 - \hat{\G}_{\hat{0}}
\hat{\G}_{\hat{1}}\hat{\G}_{\hat{11}})
\hat{\vare} = 0. \label{goodconf}
\ee
This is as far as one can go in the bulk space. However,
on the orbifold boundary planes, (\ref{z2chiral}) can be substituted in
(\ref{goodconf}) to give
\be
\frac{1}{2} (1 - \hat{\G}_{\hat{0}\hat{1}})
\hat{\vare} = 0, \ \ \ \ \  \mbox{at} \ \ \ \hat{x}^{\hat{11}} = 0, \pi \r
. \label{boundBPS} 
\ee
This expression implies that the eleven-dimensional Majorana 
spinor $\hat{\vare}$,
when restricted to the $1+1$-dimensional boundary strings, is a 
non-vanishing Majorana-Weyl spinor, as it should 
be.\footnote{When we switch to Euclidean space later in this paper, 
we must regard $\hat{\vare}$ as an eleven-dimensional Dirac spinor and
$\vare$ as a ten-dimensional Weyl spinor, since in these dimensions one
cannot impose the Majorana condition.}
We see, therefore, that this configuration preserves supercharges
consistent with the supersymmetry on the $\Z_2$ hyperplanes.
Therefore, we conclude that a configuration in which the supermembrane
is oriented parallel to the orbifold hyperplanes breaks all
supersymmetries. On the other hand, the configuration
for the open supermembrane is such that the hyperplane and membrane 
supersymmetries are compatible. Below, we will analyze the exact role
of chiral projection (\ref{boundBPS}). 

Now assume there is a bulk space five-brane oriented parallel to the
orbifold planes, and that the open supermembrane stretches between one
orbifold plane and the five-brane. The previous discussion continues
to hold on the two-dimensional boundary string, which we denote by $\del\Si_9$,
contained in the orbifold plane. What happens at the other two-dimensional
boundary string, $\del\Si_5$, embedded in $\ul{M}_6$? Clearly
(\ref{goodconf}) must remain true on $\del\Si_5$. 
It is not hard to show that this expression is compatible with the five-brane
spinor chirality constraint (\ref{5bchirlty}) and, hence, preserves
supercharges compatible with the supersymmetry on the orbifold fixed plane.
Below, we will analyze the exact role of chiral projection (\ref{goodconf})
on the supercharges of $\del\Si_5$.

We conclude that the configuration consisting of an open supermembrane
with two-dimensional boundary strings on an orbifold plane and a parallel
five-brane respectively, preserves supercharges consistent with
 the 
supersymmetry on the \HW fixed hyperplanes.

\section{$\k$-Invariant Action for Open Membranes:}

We have shown that for a supermembrane to preserve supersymmetries consistent 
with the boundary fixed planes and witha parallel oriented five-brane,
the membrane must be open. That is, it must be 
stretched between the two $\Z_2$ hyperplanes or, as we will be concerned
with in this paper, between one of the $\Z_2$ hyperplanes and the bulk space
five-brane. In this section, we want to find the action associated with such a
membrane. Action (\ref{SMaction}) is a good starting point. However, it is not
obvious that it will correspond to the desired configuration, even 
in the bulk space. For this to be the case, one needs to ask whether this 
action respects the $\Z_2$ 
symmetry of \HW theory. The answer was provided in \cite{LalLukOvr}, where 
it was concluded that, for an appropriate extension of the $\Z_2$ symmetry
to the worldvolume coordinates and similar constraints for the worldvolume
metric, the open supermembrane equations of motion are indeed $\Z_2$
covariant if the supergravity background is $\Z_2$ invariant. Therefore,
we can retain action (\ref{SMaction}). Does it suffice, however, to
completely describe the open membrane configuration? Note that the 
intersection, which we denote by $\del\Si_9$, of one end of the
open membrane with the orbifold fixed plane is a two-dimensional string 
embedded in the ten-dimensional boundary plane $M_{10}$. 
We denote by $\s^i$, $i=1,2$,
the worldsheet coordinates of this string.
Intuitively, one expects extra fields, which we generically denote by 
$\phi(\s)$, to appear on this boundary string in addition to the bulk 
fields $\hbZ^{\hbM}(\hat{\s})$. These would naturally couple to the
pullback onto the boundary string of the
background $E_8$ super-gauge fields $\BbA_{\BbM}$.
As we will see in this section, new supermembrane
fields are indeed required and form a chiral Wess-Zumino-Witten multiplet
of the $E_8$ gauge group.\footnote{This section follows closely the
original proof in \cite{Ceder}.}
Furthermore, the intersection, which we denote by $\partial\Sigma_{5}$, 
of the other end of the open supermembrane with
the bulk space five-brane is also a
$1+1$-dimensional string. However, this string is embedded in the
six-dimensional five-brane worldvolume $\ul{M}_6$. As we will see below, 
unlike the
intersection string on the orbifold plane, it is not necessary to have
extra
fields on the five-brane intersection string in addition to the bulk
fields
$\hbZ^{\hbM}(\hat{\s})$. These bulk fields 
suffice, through their derivatives along the worldsheet directions, 
to couple to the pullback onto the string worldsheet of the five-brane
super-two-form fields $\ubD_{\ubR\ubS}$. 

Let us first consider the intersection of the open membrane with the
boundary
hyperplane. As discussed previously, the supergravity theory of the
background fields exhibits both gauge and gravitational anomalies that can
only be canceled by modifying the Bianchi identity as in
(\ref{eq:burt12}). Integrating (\ref{eq:burt12}) along the 
$\hat{x}^{\hat{11}}$ direction in the neighborhood of the orbifold plane, 
and promoting the result to superspace, we find that
\be
\hbG_{\BbM\BbN\BbP\BbQ} \mid_{M_{10}} = - \frac{1}{8 \pi T_M} 
( \mbox{tr} \BbF \wedge \BbF )_{\BbM\BbN\BbP\BbQ} , 
\label{superBianchi}
\ee
where $\BbF$ is the super-field-strength of the fields $\BbA$. Note that 
we have dropped the curvature term proportional to $\mbox{tr}R \wedge R$.
The reason for this is that this term is associated, for anomaly
cancellation,
with higher dimensional terms in the eleven-dimensional supergravity
action.
However, the brane actions used here couple only to the background fields
whose dynamics are given by the usual, low dimensional supergravity
theory.
Hence, the $\mbox{tr} R \wedge R$ terms are of higher order from this
point of 
view. The reason for expressing the integrated Bianchi identity in
superspace 
is to make it compatible with the bulk part of
supermembrane action (\ref{SMaction}), which is written in terms of 
the pullbacks
of superfields $\hbE_{\hbM}^{\; \hbA}$ and $\hbC_{\hbC\hbB\hbA}$ onto the 
worldvolume. Recalling that, locally, $\hbG = d \hbC$, it follows from 
(\ref{superBianchi}) that on the orbifold plane
\be
\hbC_{\BbM\BbN\BbP} \mid_{M_{10}} =
 - \frac{1}{8 \pi T_M} \Omega_{\BbM\BbN\BbP} (\BbA), \label{CS}
\ee
where
\be
\Omega_{\BbM\BbN\BbP} (\BbA) = 3 ! 
\left( \mbox{tr} (\BbA \wedge d \BbA ) 
+ \frac{2}{3} \mbox{tr} (\BbA \wedge \BbA \wedge \BbA)
\right)_{\BbM\BbN\BbP} \label{CS2}
\ee
is the Chern-Simons three-form of the super-one-form $\BbA$.

Note that each $\BbA$ is a super-gauge-potential and, as such,
transforms under super-gauge transformations as 
\be
\d_{\BbL} \BbA_{\BbM}^a = \del_{\BbM} \BbL^a + f^{abc} 
\BbA_{\BbM}^b \BbL^c, \label{gaugtransf}
\ee
with $a,b,c=1,\ldots,248$. If we define the pullback of $\BbA$ as
\be
\BbA_i \equiv \del_i \BbZ^{\BbM} \BbA_{\BbM},
\ee
the gauge transformation in superspace (\ref{gaugtransf}) induces a gauge
transformation on the string worldsheet, which acts on the 
pullback of $\BbA$ as
\be
\d_{\BbL} \BbA_i^a = (D_i \BbL)^a = \del_i \BbL^a + f^{abc} 
\BbA_i^b \BbL^c,
\ee
where $\BbL = \BbL(\BbZ^{\BbM}(\s))$. It follows from (\ref{CS}), 
(\ref{CS2}) and (\ref{gaugtransf}) that, on the boundary fixed plane,
\be
\d_{\BbL} \hbC_{\BbM\BbN\BbP}|_{M_{10}} = - \frac{3}{4 \pi T_M} \left[
\d_{\BbL} \left( \mbox{tr} (\BbA \wedge d \BbA) 
+ \frac{2}{3} \mbox{tr} (\BbA \wedge \BbA \wedge \BbA) 
\right)_{\BbM\BbN\BbP} \right] = - \frac{3}{4 \pi T_M} 
\mbox{tr} ( \del_{[ \BbM} \BbL \del_{\BbN} \BbA_{\BbP ]} ) .
\ee
Note that, since the supergauge fields $\BbA$ live only on the orbifold
plane,
\be
\d_{\BbL} \hbC_{\BbM\BbN\BbP}=0
\label{eq:A}
\ee
everywhere else on the open supermembrane, including its intersection with
the bulk space five-brane.
Now consider the variation of the supermembrane action (\ref{SMaction})
under a super-gauge transformation. Clearly, a non-zero variation arises
from
the second term in (\ref{SMaction})
\bea
\d_{\BbL} S_{S\!M}  &=& \frac{T_M}{6} \int_{\Si} d^3 \hat{\s} \; 
\hat{\vare}^{\hi\hj \hat{k}} \del_{\hi}\hbZ^{\hbM}
\del_{\hj}\hbZ^{\hbN} \del_{\hat{k}}\hbZ^{\hbP} 
\d_{\BbL} \hbC_{\hbP\hbN\hbM} \nn \\
 &=& \frac{1}{8 \pi} \int_{\del \Si_9} d^2 \s \;
\vare^{ij} \del_i \BbZ^{\BbM} \del_j \BbZ^{\BbN}
\mbox{tr} ( \BbL \del_{\BbN} \BbA_{\BbM} ),  \label{SuG}
\eea
where we have integrated by parts. Therefore, action
(\ref{SMaction}) is not invariant under gauge transformations. This
symmetry is violated precisely at the boundary plane. It follows that to
restore
gauge invariance, one must add an appropriate boundary term
to the supermembrane action.

Before doing that, however, let us consider the transformation of the
action
$S_{S\!M} $ under a $\k$-transformation, taking into account 
the boundary expression (\ref{CS}). Note that the $\k$-transformation
acts on the super-three-form $\hbC$ as
\be
\d_{\hat{\k}} \hbC = {\cal L}_{\hat{\k}} \hbC
= i_{\hat{\k}}\, d \hbC  + (d i_{\hat{\k}})\, \hbC\, , \label{Liederiv}
\ee
where ${\cal L}_{\hat{\k}}$ is the Lie derivative in the
$\hat{\k}$-direction
and the operator $i_{\hat{\k}}$ is defined, for any super-$l$-form $\hbO$,
as
\bea
i_{\hat{\k}} \hbO &=& \frac{1}{l!} \hbO_{\hbM_{1}\cdots\hbM_{l}}
i_{\hat{\k}}
( d \hbZ^{\hbM_{l}}\wedge \cdots \wedge d \hbZ^{\hbM_1} ) \nn \\
&=& \frac{1}{(l-1)!}\hbO_{\hbM_1 \cdots \hbM_{l-1} \hat{\m}}\, (\hat{P}_+ 
\hat{\k}^{\hat{\m}} )( d \hbZ^{\hbM_{l-1}} \wedge \cdots \wedge d
\hbZ^{\hbM_1}).
\label{I} 
\eea
Importantly, we use the positive projection $\hat{P}_+$ of $\hat{\k}$, as 
defined in (\ref{kappa}), in order to remain consistent
with the previous choices of supersymmetry orientation. Varying action 
(\ref{SMaction}) under (\ref{Liederiv}), and under the full $\k$-variations
of $\hbZ$, we observe that $\k$-symmetry is also violated at the boundaries
\be
\d_{\hat{\k}} S_{S\!M}=  - \frac{1}{6} T_M \int_{\del \Si} d^2 \s \;
\vare^{ij} \del_i \BbZ^{\BbM} \del_j \BbZ^{\BbN}
\BbC_{\BbN \BbM \hat{\m}} \hat{P}_+ \hat{\k}^{\hat{\m}} 
\label{eq:B}
\ee
where $\del\Si=\del\Si_9 +\del\Si_5$ is the sum over
the two strings on the boundary of the open supermembrane. Using (\ref{CS}),
this can be written as
\bea
\d_{\hat{\k}}S_{S\!M} &=& \frac{1}{48 \pi} 
\int_{\del \Si_9} d^2 \s \;
\vare^{ij} \del_i \BbZ^{\BbM} \del_j \BbZ^{\BbN}
\Omega_{\BbN \BbM \hat{\m}} (\BbA) \hat{P}_+ \hat{\k}^{\hat{\m}} \nn \\
& & -\frac{1}{6} T_M \int_{\del \Si_{5}} d^2 \s \; \vare^{ij} \del_i 
\BbZ^{\BbM} \del_j \BbZ^{\BbN} \BbC_{\BbN \BbM \hat{\m}} \hat{P}_+ 
\hat{\k}^{\hat{\m}} \label{eq:C}
\eea
In deriving this expression, we have used the eleven-dimensional
supergravity constraints.
It proves convenient to consider, instead of this $\k$-transformation, the
modified $\k$-transformation
\be
\D_{\hat{\k}} = \d_{\hat{\k}} - \d_{\BbL_{\hat{\k}}}, \label{gkappa}
\ee
where $\d_{\BbL_{\hat{\k}}}$ is a super-gauge transformation with the
special
gauge parameter
\be
\BbL_{\hat{\k}} = i_{\hat{\k}} \BbA = 2 \BbA_{\hat{\m}} \hat{P}_+ 
\hat{\k}^{\hat{\m}}. 
\ee
Note that, although this modifies the transformation of the three-form 
$\hbC_{\BbM\BbN\BbP}$ on $\del\Si_9$, it follows from 
$(\ref{eq:A})$ that this
modified transformation is identical to the original
$\kappa$-transformation
everywhere else, including $\partial\Sigma_{5}$.
Under this transformation, the supermembrane action behaves as
\bea
\D_{\hat{\k}} S_{S\!M} &=& \frac{1}{8 \pi} \int_{\del \Si_9} d^2 \s \;
\vare^{ij} \del_i \BbZ^{\BbM} \del_j \BbZ^{\BbN}
\BbF_{\BbN \hat{\m}} \hat{P}_+ \hat{\k}^{\hat{\m}} \BbA_{\BbM} \nn \\
& & -\frac{1}{6}T_{M}\int_{\del \Si_{5}} d^2 \s \; \vare^{ij}\del_i
\BbZ^{\BbM}\del_j\BbZ^{\BbN}\BbC_{\BbN\BbM\hat{\m}}\hat{P}_+\hat{\k}^{\hat{\m}}
\label{SuK}
\eea
It is important to note that modified $\k$-invariance of $S_{SM}$ has two
independent obstructions, one on the orbifold boundary string
$\del\Si_9$ and one on the five-brane string $\del\Si_5$. Both of these,
along with the obstruction to gauge invariance on $\del\Si_9$ specified in
(\ref{SuG}),
must somehow be canceled if the theory is to be consistent. Before doing
that, it is also useful to record that the pullback of the boundary
background
field $\BbA$ transforms as
\be
\D_{\hat{\k}} \BbA_i = 2 \del_i \BbZ^{\BbM} \BbF_{\BbM \hat{\m}}
\hat{P}_+ \hat{\k}^{\hat{\m}}
\ee
under this modified $\k$-transformation, where we have used the fact that
\be
\d_{\hat{\k}} \BbA = {\cal L}_{\hat{\k}} \BbA
\ee
is the $\k$-transformation of $\BbA$, just as in (\ref{Liederiv}).

We now turn to the question of canceling both the gauge and modified
$\k$-transformations on both $\del\Si_9$ and $\del\Si_5$. We begin
with the intersection string $\del\Si_9$ of the orbifold plane
and the membrane. It was shown in \cite{Ceder} that the gauge and modified
$\k$-anomalies on $\del\Si_9$, given in (\ref{SuG}) and the first term in
(\ref{SuK}),
can be canceled if the supermembrane action is augmented to include a
chiral level one Wess-Zumino-Witten model on the orbifold boundary string 
of the membrane. The fields thus introduced will couple to the pullback of
the background field $\BbA$ at that boundary.

On $\del\Si_9$, the new fields can be written as
\be
g(\s)=e^{\phi^a (\s) T^a }, \label{e8}
\ee
where $T^a$, $a=1,\ldots,248$ are the generators of $E_8$ 
and $\phi^a (\s)$ are scalar fields that transform in the adjoint 
representation, and parametrize the group manifold, of $E_8$.
Note that $g$ is a field living on the worldsheet of the orbifold boundary
string.
The left-invariant Maurer-Cartan one-forms $\w_i(\s)$ are defined by 
\be
\w_i=g^{-1}\del_i g .
\ee
The variation of $g(\s)$ under gauge and modified $\k$-transformations
can be chosen to be
\be
\d_{\BbL} g = g \BbL , \ \ \ \ \ \ \ \ \ \ \D_{\hat{\k}} g = 0,
\label{gtransf}
\ee
where $\BbL = \BbL (\BbZ (\s))$.
The coupling of this model to the external gauge fields is 
accomplished by replacing the left-invariant Maurer-Cartan one-form
$\w_i=g^{-1}\del_i g$ by the ``gauged'' version 
\be
g^{-1} D_i \, g = \w_i - \del_i \BbZ^{\BbM} 
\BbA_{\BbM},
\ee
where $D_i$ is the
covariant derivative for the right-action of the gauge group. 

An action that is gauge- and $\k$-invariant on the membrane bulk space
worldvolume and on $\del\Si_9$, but not yet on the five-brane boundary
string, can be obtained by adding to the bulk action $S_{SM}$ given in  
(\ref{SMaction}) the Wess-Zumino-Witten action
\bea
S_{W\!Z\!W}  &=& \frac{1}{8 \pi} \int_{\del \Si_{10}} d^2 \s \;
\mbox{tr} [\frac{1}{2} \sqrt{-g} g^{ij} ( \w_i - \del_i \BbZ^{\BbM} 
\BbA_{\BbM} ) \cdot ( \w_j - \del_j \BbZ^{\BbN} 
\BbA_{\BbN} )  + \vare^{ij} \del_j \BbZ^{\BbM} 
\w_i \BbA_{\BbM} ] \nn \\
& & - \frac{1}{24 \pi} \int_{{\cal{B}}} d^3 \hat{\s} \; 
\hat{\vare}^{\hi\hj \hat{k}} \Omega_{\hat{k}\hj\hi} (\hat{\w}) ,
\label{WZWaction}
\eea
where
\be
\Omega_{\hat{k}\hj\hi}(\hat{\w})=
\mbox{tr}(\hat{\w}\wedge\hat{\w}\wedge\hat{\w})_{\hat{k}\hj\hi}\label{wzw3f}
\ee
and we use $\hat{\vare}_{012} = + 1$.
The first term in (\ref{WZWaction}) describes the kinetic 
energy for the scalar fields $\phi^a (\s)$ and their interactions with the
pullback of the super-gauge potential $\BbA$. The second term is the 
integral, over the three-ball ${\cal{B}}$ with boundary $\del\Si_9$, of
the Wess-Zumino-Witten three-form, constructed in (\ref{wzw3f}) from a
one-form $\hat{\w} = \hat{g}^{-1} d \hat{g}$, where $\hat{g} : {\cal{B}}
\to
E_8$. The map $\hat{g}$ must satisfy
\be
\hat{g} \mid_{\del \Si_9} = g,  
\label{boundg}
\ee
but is otherwise unspecified. That such a $\hat{g}$ exists was shown
in \cite{Wnab}. Note that we have implicitly assumed that
\be
\del\Si_9 = \BbC\BbP^1 = S^2 , \label{9issphere}
\ee
as we will do later in this paper when computing the superpotential.
It is straightforward to demonstrate that
the variation of $S_{W\!Z\!W} $ under both gauge 
and local modified $\k$-transformations, $\d_{\BbL}$ and  $\D_{\hat{\k}}$ 
respectively, exactly cancels the variations of the
bulk action $S_{S\!M} $ given in (\ref{SuG}) and the first term in
(\ref{SuK}), provided we choose the parameter $\hat{\k}$ on $\del\Si_9$ 
to obey
\be
P_- \hat{\k}  = 0 , 
\ee
where the projection operators $P_{\pm}$ are defined as
\be
P_{\pm} \equiv \frac{1}{2} (1 \pm \frac{1}{2\sqrt{-\det g_{ij}}} 
\vare^{ij} \Pi_i^{\, A} \Pi_j^{\, B} \G_{AB} ) 
\ee
Note that this is consistent with (\ref{boundBPS}). On the orbifold
boundary
string we can denote $\hat{\k}$ by $\k$. In proving this cancellation, it
is necessary to use the super-Yang-Mills constraints on the boundary hyperplane.
We conclude that by adding $S_{W\!Z\!W}$ to the action, we have completely
canceled the gauge anomaly (\ref{SuG}) and the $\del\Si_9$ term in the modified
$\k$-anomaly (\ref{SuK}). We now turn to the question of canceling the
remaining obstruction to modified $\k$-invariance, namely, the
$\del\Si_5$ term in (\ref{SuK}).

Before doing this, it is necessary to discuss the embedding coordinates 
of the boundary string on the five-brane. First note that, when restricted 
to the intersection between the membrane and the five-brane,
\be
\BbZ^{\BbM} (\s) = \ubY^{\BbM}(\s). \label{ZisY}
\ee
Furthermore, it follows from the static gauge choice (\ref{5bconfig}) of the 
five-brane that all supercoordinates $\ubY^{\BbM}(\s)$ vanish except for those
given in terms of the intrinsic worldvolume supercoordinates as
\be
\ubY^{\ubR}(\s) = (\xi^{\ul{r}}(\s),\T^{\ul{\m}}(\s)) ,
\label{superY}
\ee
where $\T^{\ul{\m}}$ is a 
16-component
spinor of (2,0)-supersymmetry in six-dimensions. In terms of the
supercoordinates in (\ref{superY}), the five-brane tensor supermultiplet
(\ref{5bsupmult}) can be expressed as a super-two-form $\ubD_{\ubR\ubS}$
satisfying
\be
\ubD_{\ul{r}\ul{s}} \mid_{\T^{\ul{\m}} = 0} = \ul{D}_{\ul{r}\ul{s}}
\ee
and another constraint that we will specify below.
It is useful to note that the second term in (\ref{SuK}) can now be written
as
\be
-\frac{1}{6}T_{M}\int_{\del \Si_{5}} d^2 \s \; \vare^{ij}\del_i
\ubY^{\ubR}\del_j\ubY^{\ubS} \ubC_{\ubS\ubR \ul{\m}} P_+
\k^{\ul{\m}}, \label{newterm}
\ee
where we have used the fact that the string can only move within
the five-brane worldvolume and, therefore, $\uh{\k}$ must be projected
as in (\ref{5bchirlty}).

It was shown in \cite{chusezgin} that the $\del\Si_{5}$ term in (\ref{SuK}), that is,
the obstruction to $\k$-invariance on the intersection string of the
membrane and five-brane, can be canceled if one adds to the action
$S_{S\!M}+S_{W\!Z\!W}$ another term, supported only on $\del\Si_{5}$,
given by
\begin{equation}
S_{5}=\frac{1}{6} T_M \int_{\del\Si_{5}} d^2 \s \vare^{ij} \ubD_{ij}
\label{eq:alpha}
\end{equation}
where $i,j=0,1$ and 
\begin{equation}
\ubD_{ij}=del_i \ubY^{\ubR} \del_j \ubY^{\ubS} \ubD_{\ubS\ubR}
\label{eq:beta}
\end{equation}
is the pullback of the five-brane worldvolume super-two-form
$\ubD_{\ubR\ubS}$ onto the boundary string $\del\Si_{5}$. It is
important to note that, unlike the case of the Wess-Zumino-Witten action
(\ref{WZWaction}), it is not necessary to introduce any new dynamical
degrees
of freedom on $\partial\Sigma_{5}$ in order to couple this string to the
background five-brane worldvolume two-form superfield. Clearly, this piece
of the action does not involve the
$E_{8}$ supergauge fields of the orbifold boundary in any way and, hence,
the gauge anomaly continues to vanish. However, $\ubD_{\ubR\ubS}$ does have
a non-trivial transformation under modified $\k$-transformations. This is
given by
\begin{equation}
\Delta_{\k}\ubD=\delta_{\k}\ubD=
i_{\k}d\ubD+(di_{\k})\ubD
\label{eq:gamma}
\end{equation}
where the action of $i_{\k}$ on any super-l-form is defined in
(\ref{I}). 
Varying $S_{5}$ under the modified $\k$-transformation, we find that
\begin{equation}
\Delta_{\hat{\kappa}}S_{5}=\frac{1}{6}T_{M} \int_{\del \Si_{5}} d^2 \s
\vare^{ij}\del_i \ubY^{\ubR} \del_j \ubY^{\ubS}(d\ubD)_
{\ubS\ubR\uh{\m}} P_+ \k^{\ul{\m}} .
\label{eq:delta}
\end{equation}
Using (\ref{newterm}), we can add this variation to the $\partial\Sigma_{5}$ 
term in (\ref{SuK}) giving
\begin{equation}
\Delta_{\hat{\k}} S_{SM} \mid_{\del\Si_{5}} + \Delta_{\hat{\k}} S_{5} =
\frac{1}{6} T_{M} \int_{\del \Si_{5}} d^2 \s \vare^{ij} \del_i \ubY^{\ubR} \del_j 
\ubY^{\ubS} \ubH_{\ubS\ubR \ul{\m}} P_+ \k^{\ul{\m}},
\label{eq:mew}
\end{equation}
where
\begin{equation}
\ubH=d\ubD-\ubC .
\label{help}
\end{equation}
This variation will vanish and modified $\kappa$-invariance will be restored
if and only if we impose the constraint
\be
\ubH_{\ubS\ubR \ul{\m}} = 0 . \label{HrRSiszero}
\ee
This constraint reduces
the large reducible multiplet in $\ubD_{\ubR\ubS}$ to the irreducible
tensor supermultiplet specified in (\ref{5bsupmult}).

Thus far, we have demonstrated that the combined action for the open
supermembrane with boundary strings
\begin{equation}
S_{O\!M}=S_{S\!M}+S_{W\!Z\!W}+S_{5}
\label{eq:EE}
\end{equation}
is invariant under both $E_{8}$ gauge transformations on $\del\Si_9$
and modified $\k$-transformations everywhere. We have, however, yet to
check that this action is invariant under the Abelian transformations
\begin{equation}
\d_{\L}\ubD =d \ul{\L}
\label{eq:AA}
\end{equation}
of the super two-form on $\partial\Sigma_{5}$. As in the case of $E_{8}$ 
supergauge transformations, there are two potential sources of
$\d_{\L}$ anomalies. First, integrate the modified Bianchi identity
(\ref{eq:burt12}) along the $\hat{x}^{11}$ direction in the neighborhood
of
the five-brane. We find that
\begin{equation}
\hbG_{\BbM\BbN\BbP\BbQ} \mid_{\ul{M}_6}=0
\label{eq:BB}
\end{equation}
Exactly as for the $\mbox{tr}R\wedge R$ term at the orbifold boundary
$M_{10}$, we
have dropped the contribution from the five-brane source $J_{5}$ on the
right
hand side of (\ref{eq:burt12}). Again, the reason for doing this is that
this term
is associated, for anomaly cancellation, with higher dimensional terms in
the
eleven-dimensional supergravity action. However, the five-brane action
used
here couples only to background fields associated with the usual, low
dimension terms in this supergravity theory. Hence, the $J_{5}$ terms are
higher order from this point of view. It follows from (\ref{eq:BB}) that
\begin{equation}
\hbC_{\BbM\BbN\BbP}\mid_{M_{5}}=(d\tilde{\BbD})_{\BbM\BbN\BbP}
\label{eq:CC}
\end{equation}
where $\tilde{\BbD}$ is some super-two-form on $\ul{M}_6$ unrelated to
$\ubD$. Note that this expression
is consistent with the constraint equation (\ref{HrRSiszero}). 
It follows that 
\be
\d_{\L} \BbC_{\BbM\BbN\BbP} \mid_{\ul{M}_6} = 0
\ee
and, hence,
$S_{S\!M}$ is invariant. Second, note that since the variation of
its integrand is a total
divergence, $S_{5}$ is invariant under Abelian transformation
(\ref{eq:AA}). Combining these results, we conclude that
\begin{equation}
\d_{\L}S_{O\!M}=0 .
\label{eq:DD}
\end{equation}

\section{Low-Energy Limit and the Heterotic Superstring:}

In this paper, we are interested in obtaining an effective
four-dimensional
theory with $N=1$ supersymmetry. In particular, we want to compute 
non-perturbative corrections to the superpotential of the theory. These
corrections arise from the non-perturbative interaction between the
background
and the open supermembrane embedded in it. The total action of this theory
is 
\begin{equation}
S_{Total}=S_{H\!W}+S_{O\!M}=
(S_{S\!G}+S_{Y\!M})+(S_{S\!M}+S_{W\!Z\!W}+S_{5})
\label{eq:FF}
\end{equation}
where $S_{S\!G}$, $S_{Y\!M}$ can be found in \cite{LOPR} and $S_{S\!M}$,
$S_{W\!Z\!W}$ and
$S_{5}$ are given in (\ref{SMaction}), (\ref{WZWaction}) and 
(\ref{eq:alpha}) respectively. In addition to compactifying on
$S^1/\Z_2$, which takes eleven-dimensional supergravity to the \HW theory, 
there must be a second dimensional reduction on a real
six-dimensional manifold. This space, which reduces the theory from ten-
to
four-dimensions on each orbifold boundary plane, and from eleven- to
five-dimensions in the bulk space, is taken to be a Calabi-Yau threefold,
denoted $CY_3$. A Calabi-Yau space is chosen since such a configuration
will
preserve $N=1$ supersymmetry in four-dimensions. That is, we now consider
$M$-theory, open supermembranes and five-branes on the geometrical
background
\begin{equation}
M_{11}=R_{4} \times CY_{3} \times S^1/\Z_2
\label{eq:ef}
\end{equation}
where $R_{4}$ is four-dimensional, flat space.

It is essential that this theory be Lorentz invariant in four-dimensions.
Consider a five-brane located in the bulk space and oriented parallel to
the
orbifold fixed planes. It is clear that to maintain Lorentz invariance,
the
manifold of the five-brane must be of the form
\begin{equation}
\ul{M}_6 = R_4 \times \CC
\label{eq:GG}
\end{equation}
where $\CC$ is a real two-dimensional surface with the property that
\begin{equation}
\CC \subset CY_3
\label{eq:HH}
\end{equation}
It was shown in \cite{B31} that, in order to preserve $N=1$ four-dimensional
supersymmetry on $R_4$, $\CC$ must be a holomorphic curve in
$CY_3$.
Now consider an open supermembrane stretched between one orbifold plane
and 
the bulk space five-brane. Any such membrane must have an embedding
geometry
given by
\begin{equation}
\Sigma=\tilde{\cal{C}} \times I
\label{eq:II}
\end{equation}
where $\tilde{\cal{C}}$ is a real, two-dimensional surface and $I\subset
S^1/\Z_2$ is the interval in the orbifold direction between the orbifold
plane and the five-brane. Clearly, the requirement of four-dimensional
Lorentz invariance implies that
\begin{equation}
\tilde{\cal{C}}\subset CY_{3}
\label{eq:JJ}
\end{equation}
Since $CY_{3}$ is purely space-like, it follows that we must, henceforth,
use
the Euclidean version of supermembrane theory.\footnote{Another 
reason to Euclideanize the theory is that,
in this paper, we will perform the calculation of quantum corrections using 
the path-integral formalism.}
It was shown in \cite{LOPR} that, in order to preserve $N=1$
supersymmetry in four-dimensions, it is necessary to choose
$\tilde{\CC}$ to be a holomorphic curve in $CY_3$. Clearly, since $\Si$ has a
boundary in $\ul{M}_6$ we must have
\begin{equation}
\Sigma={\cal{C}} \times I
\label{eq:KK}
\end{equation}

In this section, we take the limit as the radius $\r$ of $S^1$ becomes
small and explicitly compute the open supermembrane theory in this limit. 
The result will be the heterotic superstring, coupled to
one $E_{8}$ gauge background and to a Neveu-Schwarz five-brane, embedded
in the ten-dimensional space
\be
M_{10} = R_4 \times CY_3,
\ee
and wrapped around a holomorphic curve $\CC \subset CY_3$.

We begin by rewriting the action (\ref{eq:EE}) for an open supermembrane 
with boundary strings on one orbifold plane and a bulk space five-brane as
\bea
S_{O\!M}  &=& T_M \int_{\Si} d^3 \hat{\s} ( 
\sqrt{\det \hat{\Pi}_{\hi}^{\, \hat{A}} \hat{\Pi}_{\hj }^{\, \hat{B}}
\h_{\hat{A}\hat{B}}} - \frac{i}{6} 
\hat{\vare}^{\hi\hj \hat{k}} \hat{\Pi}_{\hi}^{\, \hbA}
\hat{\Pi}_{\hj}^{\, \hbB} \hat{\Pi}_{\hat{k}}^{\, \hbC} 
\hbC_{\hbC\hbB\hbA} ) \nn \\
& & - \frac{1}{8 \pi}  \int_{\del \Si_{9}} 
 d^2 \s \; \ \mbox{tr}[\frac{1}{2} \sqrt{g}g^{ij} ( \w_i - \BbA_i )
 \cdot ( \w_j - 
\BbA_j ) + i \vare^{ij} \w_i \BbA_j ]\nn \\
& &  +\frac{1}{24\pi} \int_{{\cal{B}}}{d^3 \hat{\sigma} i\vare
^{\hat{i}\hat{j}\hat{k}}\Omega_{\hat{k}\hat{j}\hat{i}}(\hat{\omega})}
-\frac{1}{6} T_{M} \int_{\del\Si_{5}} d^2 \s i \vare^{ij} \ubD_{ij} . 
\label{OMaction2}
\eea
An $i$ appears multiplying the epsilon symbols because we are in Euclidean space.
Furthermore, it is important to note that the requirement that we
work in Euclidean space changes the sign of each term in (\ref{OMaction2})
relative to the Minkowski signature action given by 
(\ref{WZWaction}) and (\ref{eq:alpha}).
The boundary terms describe the gauged chiral Wess-Zumino-Witten model on
the
orbifold string and the coupling to the super-two-form on the
five-brane string. Since they are defined only on the boundary, 
they are not affected by the compactification on $S^1/\Z_2$. As for the
bulk action, we identify 
\begin{equation}
\hat{X}^{\hat{11}} = \hat{\s}^{\hat{2}} 
\label{eq:LL}
\end{equation}
and for all remaining fields keep only the 
dependence on $\hat{\s}^{\hat{0}}, \hat{\s}^{\hat{1}}$. The explicit
reduction of the bulk action was carried out in \cite{LOPR}, to which we refer the
reader. Here, we will simply state the result. We find that
the first part of action (\ref{OMaction2}) reduces in the small $\r$
limit to the string action
\be
S_S  = T_S \frac{{Y}}{\pi\rho}\int_{\CC} d^2 \s ( \phi 
\sqrt{\det \Pi_i^{\, A} \Pi_j^{\, B}
\h_{AB}} - \frac{i}{2} \vare^{ij} \Pi_i^{\, \BbA}
\Pi_j^{\, \BbB} \BbB_{\BbB\BbA} ),      \label{TypeIAction}
\ee
where 
\be
T_S = T_M \pi \rho \equiv (2 \pi \a')^{-1} \label{tsistm}
\ee
is the string tension of mass dimension two, super-two-form $\BbB_{\BbB\BbA}$ 
and dilaton superfield $\phi$ are defined below and ${Y}$ is the location
coordinate of the five-brane in the $S^{1}/\BbZ_{2}$ orbifold interval.
This coordinate is chosen so that when ${Y} \to 0$, 
the length of the open membrane shrinks to zero. This important factor
arises from the fact that, by assumption, no fields depend on intrinsic
coordinate $\hat{\sigma}^{\hat{2}}$ and that
\begin{equation}
\int{d^{3} \hat{\s}} = \int_0^{{Y}} d \hat{\s}^{\hat{2}}
\int d^2 \s = {Y} \int d^2 \s   
\label{eq:MM}
\end{equation}

Before we can write the total action for the open supermembrane
compactified
on $S^1/\Z_2$, we must discuss the boundary terms in (\ref{OMaction2}).
In the limit that the radius $\r$ of $S^1$ shrinks to zero,
the orbifold fixed plane and the five-brane coincide. Generically,
the two different boundaries of the supermembrane need not be identified.
However, since our supersymmetric embedding Ansatz (\ref{eq:LL}) assumes all 
quantities
to be independent of the orbifold coordinate, the two boundary strings
coincide as the zero radius limit is taken. This has further implications
beyond the fact that, at low energy, we are dealing with a single string.
To see this, begin by considering the full orbifold before taking the
small $\r$ limit and before compactifying on $CY_3$. Note that,
prior to the embedding Ansatz, the membrane boundary on the orbifold
fixed plane, $\del\Si_9$, can be any two-dimensional subset of $M_{10}$.
However, Ansatz (\ref{eq:LL}) implies that
\be
\del\Si_9 \subset \ul{M}_6 \subset M_{10},
\ee
where $\ul{M}_6$ is the induced embedding of the five-brane manifold into 
$M_{10}$. This constraint limits the bosonic target space coordinates of 
$\del\Si_9$ to lie in a six-dimensional submanifold of $M_{10}$ and will have
important implications that will be discussed later in this paper. Furthermore,
the restriction of $\del\Si_9$ to $\ul{M}_6\subset M_{10}$ implies that the
five-brane chiral constraint (\ref{5bchirlty}) now applies to the 
supercharges on $\del\Si_9$, in addition to the $\Z_2$ induced chiral 
constraint (\ref{z2chiral}).
This reduces from 16 to 8 the number of preserved supercharges on $\del\Si_9$.
The embedding Ansatz, however, prior to taking the small $\r$ limit has no
effect on the coordinates, bosonic or fermionic, of $\del\Si_5$.

Now take the limit that $\r\to 0$. In this limit, there is no change on
$\del\Si_9$. However, in the small radius limit, the $\Z_2$ projection
(\ref{z2chiral}) applies to supercharges on $\del\Si_5$ in addition to the
chiral constraint (\ref{5bchirlty}), reducing them from 16 to 8.
They are given by exactly the same supercharges as
on $\del\Si_9$. The small $\r$ limit does not affect the bosonic coordinates
of $\del\Si_5$ which, by definition, will satisfy $\del\Si_5\subset
\ul{M}_6\subset M_{10}$.

Note that our analysis of the supercharges of both $\del\Si_9$ and 
$\del\Si_5$, in the small $\r$ limit, remains incomplete. As discussed
previously, prior to taking $\r$ small, the supercharges on $\del\Si_9$
are further restricted by chiral constraint (\ref{boundBPS}) and those on
$\del\Si_5$ by chiral constraint (\ref{goodconf}). In the $\r\to 0$ limit, 
these constraints become identical. This constraint further reduces the number
of supercharges from 8 to 4. We conclude that,
as $\r\to 0$, the boundary strings coincide so that
\be
\CC = \del\Si_9 = \del\Si_5 \label{CCin5and9}
\ee
and satisfy
\be
\CC \subset \ul{M}_6 \subset M_{10}
\ee
with four preserved supercharges. These restrictions are important, as we
will see below. Putting everything together, we find that the resulting
action is
\bea
S_{\CC} &=& T_S \frac{{Y}}{\pi\rho}\int_{\CC} d^2 \s ( \phi 
\sqrt{\det \Pi_i^{\, A} \Pi_j^{\, B}
\h_{AB}} - \frac{i}{2} \vare^{ij} \Pi_i^{\, \BbA}
\Pi_j^{\, \BbB} \BbB_{\BbB\BbA} ) \nn \\
& & - \frac{1}{8 \pi} \int_{\CC} 
 d^2 \s \; \mbox{tr}[\frac{1}{2} \sqrt{g}g^{ij} ( \w_i - \BbA_i )
 \cdot ( \w_j - 
\BbA_j ) + i \vare^{ij} \w_i \BbA_j ]\nn \\
& & +\frac{1}{24\pi}\int_{\cal{B}}  d^3\hat{\s}
i\hat{\vare}^{\hi\hj\hat{k}}\Omega_{\hat{k}\hj\hi}
(\hat{\omega}) -\frac{1}{6} T_S\int_{\CC} d^2 \s i \vare^{ij} \ubD_{ij},
\label{HSaction}
\eea
where
\be
\Pi_i^{\, \BbA} = \del_i \BbZ^{\BbM} \BbE_{\BbM}^{\; \BbA}.
\ee
and
\begin{equation}
\BbB_{\BbM\BbN}=\hbC_{\BbM\BbN\hat{11}}, \qquad
\phi=\hbE_{\hat{11}}^{\;\hat{11}} .
\label{eq:one}
\end{equation}
Note that in the last term of (\ref{HSaction}) we have used (\ref{tsistm})
and absorbed a factor of $1/\pi\rho$ into the definition of the
superfield $\ubD$ so that it now has mass dimension zero.
For ease of notation, we have written (\ref{HSaction}) in terms of the
ten-dimensional superembedding coordinates
\be
\BbZ^{\BbM} = (X^M , \T^{\m}),
\ee
where spinor $\T$ satisfies the Weyl chirality constraint
\be
\frac{1}{2}(1- \G_{11})\T = 0 . \label{weylcond}
\ee
However, as we have just discussed, the superembedding is to be considered
further restricted to
\be
\BbZ^{\ubR} = \ubY^{\ubR} = (\xi^{\ul{r}},\T^{\ul{\m}}) ,
\ee
where $\ul{r}=0,1,\ldots,5$ and $\T$ satisfies the additional, gauge-fixing 
conditions that
\be
\frac{1}{2}(1+i \G_{012345})\T = 0 , \ \ \ \ \ \ \ \ \ \ \ \ \ \ \
\frac{1}{2}(1+i \G_{01})\T = 0 . \label{addcond}
\ee
The chiral projections in (\ref{weylcond}) and (\ref{addcond}) reduce the
number of independent components of spinor $\T$ to four.
We note in passing that the dilaton superfield $\phi$ satisfies
\begin{equation}
\hat{g}_{\hat{11}\hat{11}}=\phi^{2}.
\label{eq:two}
\end{equation}
This expression will be useful in the next section when discussing low
energy
moduli fields. We recognize the action (\ref{HSaction})
as that of the heterotic superstring coupled to one $E_{8}$ gauge
background,
a Neveu-Schwarz five-brane and wrapped on a holomorphic 
curve $\CC \subset CY_3$. In this paper, the curve $\CC$ is restricted to
\be
\CC = \BbC \BbP^1 = S^2 .
\ee
This follows from expressions (\ref{9issphere}) and (\ref{CCin5and9}).

\section{Superpotential in 4D Effective Field Theory:}

It is essential when constructing superpotentials to have a detailed 
understanding of all the moduli in five-dimensional
heterotic $M$-theory. Furthermore, we must know
explicitly how they combine to form the moduli of the four-dimensional
low-energy theory. The
compactification of \HW theory to heterotic $M$-theory on a Calabi-Yau
threefold with
$G$-flux, but without bulk five-branes, was carried out in \cite{B34,B32}, and
reviewed in
\cite{burtlec}. The further compactification of this theory on $S^1/\Z_2$, arriving at
the $N=1$ sypersymmetric action of the effective four-dimensional theory
was
presented originally in \cite{B32} and, again, was reviewed in \cite{burtlec}. 
We refer the reader
to these papers for all necessary details. Here, we discuss only those
relevant
moduli not reviewed in \cite{burtlec}, namely, the moduli associated with the
translation
of the bulk-space five-brane. We emphasize that, throughout this paper, 
we take the bosonic components of all superfields to be of dimension zero, 
both in five-dimensional heterotic
$M$-theory and in the associated four-dimensional effective theory.

First, consider the compactification from \HW theory to heterotic $M$-theory.
This compactification is carried out as follows. Consider the metric
\be
ds_{11}^2=V^{-2 / 3} g_{\hat{u}\hat{v}} d\hat{y}^{\hat{u}}
d\hat{y}^{\hat{v}} +
g_{\breve{U}\breve{V}} d \breve{y}^{\breve{U}} d \breve{y}^{\breve{V}},
\label{cymetric}
\ee
where $\hat{y}^{\hat{u}}$, $\hat{u}=2,3,4,5,11$ are the coordinates of the
five-dimensional bulk space of heterotic $M$-theory,
$\breve{y}^{\breve{U}}$,
$\breve{U}=0,1,6,7,8,9$ are the Calabi-Yau coordinates and
$g_{\breve{U}\breve{V}}$ is the metric on the Calabi-Yau space $CY_3$. 
The factor $V^{-2/3}$ in (\ref{cymetric})
has been chosen so that metric $g_{\hat{u}\hat{v}}$ is the
five-dimensional Einstein frame metric.
The Calabi-Yau volume modulus $V = V(y^{\hat{u}})$ is
defined by
\be
V = \frac{1}{v} \int_{CY_3} \sqrt{\breve{g}} ,
\ee
where $\breve{g}$ is the determinant of the Calabi-Yau metric 
$g_{\breve{U}\breve{V}}$ and $v$ is a dimensionful parameter necessary to
make $V$ dimensionless. 

These fields all must be the bosonic components of specific $N=1$ 
supermultiplets in five-dimensions. These supermultiplets are easily
identified as follows.

\noindent 1. Supergravity: the bosonic part of this supermultiplet is
\be
(g_{\hat{u}\hat{v}},\cA_{\hat{u}},\ldots).
\ee
This accounts for $g_{\hat{u}\hat{v}}$. The origin of the graviphoton
component $\cA_{\hat{u}}$ was discussed in \cite{B32}. 

\noindent 2. Universal Hypermultiplet: the bosonic part of this
supermultiplet is
\be
(V, C_{\hat{u}\hat{v}\hat{w}},\xi,\ldots),
\ee
which accounts for the Calabi-Yau volume modulus $V$. The remaining zero-modes
components were discussed in \cite{B32}. Having identified the appropriate $N=1$, 
five-dimensional superfields, one can
read off the zero-mode fermion spectrum to be precisely those fermions
that complete these supermultiplets.

Thus far, we have not said anything about the bulk space five-brane. As
discussed in Section 2, after fixing the $\k$-gauge the worldvolume 
theory of the
five-brane exhibits $(2,0)$-supersymmetry. The worldvolume fields of the
five-brane form a tensor supermultiplet.

\noindent 3. Tensor Supermultiplet: The complete supermultiplet is
\be
(\ul{D}_{\ul{r}\ul{s}},\uh{Y}^{\hat{p}},\chi), \ \ \ \ \ \ \ \ \ \ \ \ \
\hat{p} = 6,\ldots,9,11,
\ee
where the field-strength of $\ul{D}_{\ul{r}\ul{s}}$ is anti-self-dual, 
there are five scalars $\uh{Y}^{\hat{p}}$ and $\chi$ are
the associated fermions. For a five-brane oriented parallel to the orbifold
fixed planes, four of the scalars $\ul{Y}^p$, $p=6,\ldots,9$ are moduli in
the Calabi-Yau direction and we can ignore them. The fifth scalar
$Y^{\hat{11}}$, 
which we now simply refer to as ${Y}$, is the translational mode 
of the five-brane in
the orbifold direction and is of principal interest in this paper. All of
these fields are functions of the six worldvolume coordinates
$\xi^{\ul{r}}$, $\ul{r}=0,1,\ldots,5$. 

We now move to the discussion of the compactification of heterotic
$M$-theory in five-dimensions to the effective $N=1$ supersymmetric theory
in four-dimensions. This compactification, without the five-brane, was
carried out in detail in \cite{B32} and reviewed in
\cite{burtlec}. Here, we simply state the relevant four-dimensional zero-modes and
their exact
relationship to the five-dimensional moduli of heterotic $M$-theory.
The bulk space zero-modes coincide with the $\Z_2$-even fields. One
finds that the metric is
\be
ds_5^2 = R^{-1} g_{uv} dy^u dy^v + R^{2} (d y^{\hat{11}})^2 ,
\label{5dmetric}
\ee
where $g_{uv}$ is the four-dimensional metric, $R = R(y^u)$ is the
volume modulus of $S^1/\Z_2$ and $y^u$, $u=2,3,4,5$ are the four-dimensional
coordinates. The Calabi-Yau volume modulus reduces to
\be
V=V(y^u).
\ee
It is conventional to incorporate this field into the complex dilaton
$S$ as
\be
S = V + i \sqrt{2} \s 
\label{sandt}
\ee
where scalar field $\s$ was discussed in \cite{B32}. Furthermore, there
are an additional $h^{1,1}$ $(1,1)$-moduli, denoted by $T^{I}$, which
arise in
the context of superpotentials and were defined in detail in \cite{B32}.
Of importance in this paper is a particular linear combination of
these $(1,1)$-moduli, which we denote by $\cT$.
Modulus $\cT$ is related to the
$(1,1)$-moduli $T^I$ as follows. Recall that the cohomology group $H^{(1,1)}$
on $CY_3$ has a basis of harmonic $(1,1)$-forms $\w_I$, $I=1,\ldots,h^{1,1}$.
These are naturally dual to a basis $\CC_I$, $I=1,\dots,h^{1,1}$ of curves
in $H_{(1,1)}$ where
\be
\frac{1}{v_{\CC}} \int_{\CC_I} \w_J = \d_{IJ} . \label{innerprod}
\ee
We have introduced a parameter $v_{\CC}$ of mass dimension minus two to make 
the integral dimensionless. Parameter $v_{\CC}$ can be taken to
be the volume of curve $\CC$.
Any holomorphic curve can be expressed as a linear combination of the $\CC_I$
curves. For example, the curve $\CC$ around which our heterotic string is
wrapped can be written
\be
\CC = \sum_{I=1}^{h^{1,1}} c_I \CC_I \label{lincomb}
\ee
for some complex coefficients $c_I$, $I=1,\ldots,h^{1,1}$. The dual to this
expression is the harmonic $(1,1)$-form
\be
\w_{\CC} = \frac{1}{(\sum_{K=1}^{h^{1,1}}c_K^2)} \sum_{I=1}^{h^{1,1}}c_I \w_I ,
\ee
where
\be
\frac{1}{v_{\CC}} \int_{\CC} \w_{\CC} = 1 . \label{CCandw}
\ee
This form can be extended to a basis of $H^{(1,1)}$. Denote the remaining
$h^{1,1}-1$ basis forms by $\w'_i$, with the property
\be
\frac{1}{v_{\CC}} \int_{\CC} \w'_i = 0 . \label{w'isortho}
\ee
Now, note from the discussion in \cite{LOPR} that
\be
R V^{-1/3} \w = \sum_{I=1}^{h^{1,1}} \mbox{Re} T^I \w_I ,
\ee
where $\w$ is the K\"ahler form on $CY_3$. Similarly, 
one can define $\mbox{Re} \cT$ by
\be
R V^{-1/3} \w = \mbox{Re} \cT \w_{\CC} + \sum_{i=1}^{h^{1,1}-1} 
\b^i \w'_i . \label{expan}
\ee
Equating these two expressions and integrating over $\CC$ using
(\ref{innerprod}), (\ref{lincomb}), (\ref{CCandw}) and (\ref{w'isortho}), 
we find that
\be
\mbox{Re} \cT = \sum_{I=1}^{h^{1,1}} c_I \mbox{Re}T^I .
\label{reTao}
\ee
Furthermore, from the discussion in \cite{LOPR} we note that
\be
B = \sum_{I=1}^{h^{1,1}} \mbox{Im}T^I \w_I ,
\ee
where $B_{m\bn}=\hat{C}_{m\bn\hat{11}}$ is the bosonic component of
superfield $\BbB_{\BbM\BbN}$ defined in (\ref{eq:one}). Similarly, one
can define $\mbox{Im}\cT$ by
\be
B = \mbox{Im} \cT \w_{\CC} + \sum_{i=1}^{h^{1,1}-1} \g^i \w'_i .
\label{Bisimtao}
\ee
Integrating these two expressions over $\CC$ using (\ref{innerprod}),
(\ref{lincomb}), (\ref{CCandw}) and (\ref{w'isortho}), we find that
\be
\mbox{Im} \cT = \sum_{I=1}^{h^{1,1}} c_I \mbox{Im} T^I .
\label{imTao}
\ee
Putting equations (\ref{reTao}) and (\ref{imTao}) together,
we conclude that
\be
\cT = \sum_{I=1}^{h^{1,1}} c_I T^I .
\ee

The exact form of the four-dimensional $N=1$ translational supermultiplet of
the five-brane has to be carefully discussed at this point. 
It was shown in \cite{B31} that, when a
five-brane is compactified to four-dimensions on a holomorphic curve $\CC$
of genus $g$, there are two types of $N=1$ zero-mode supermultiplets that
arise. First, there are $g$ Abelian vector superfields. Since we are concerned
with superpotentials in this paper, these superfields are not of interest to us
and we will mention them no further. The second type of multiplet that arises
is associated with the translational scalar mode, now reduced to
\be
{Y} = {Y} (y^u).
\ee
In addition, one must consider the four-dimensional modulus associated with 
the two-form $\ul{D}_{\ul{r}\ul{s}}$. This is found by expanding
\be
\ul{D} = 3 a \w_{\CC} , \label{disa}
\ee
where $a=a(y^u)$. It was shown in \cite{Derend}, in an entirely different
context, that the $N=1$ translational supermultiplet of the five-brane
is a chiral multiplet whose bosonic component is given by
\be
{\bf Y} = \frac{{Y}}{\pi\rho} \mbox{Re} \cT +
i (a + \frac{{Y}}{\pi\rho} \mbox{Im} \cT). \label{bigY}
\ee
The divisor $\pi\rho$ renders ${Y}/\pi\rho$ and, hence, 
${\bf Y}$ dimensionless.

It is then easily seen that these modes form the following
four-dimensional, $N=1$ supermultiplets.

\noindent 1. Supergravity: the full supermultiplet is
\be
(g_{uv},\psi_u^{\a}),
\ee
where $\psi_u^{\a}$ is the gravitino.

\noindent 2. Dilaton and T-Moduli Chiral Supermultiplets: the full
multiplets
are
\be
(S,\l_S), \ \ \ \ \ \ \ \ \ \ \ \ \ (T^I, \l_T^I), \label{STmoduli}
\ee
where $I=1,\ldots,h^{1,1}$ and $\l_S$, $\l_T^I$ are the 
dilatino and T-modulinos, respectively. In particular, the $\cT$ modulus
is the lowest component of chiral superfield
\be
(\cT,\l_{\cT})
\ee

\noindent 3. Five-Brane Translation Chiral Supermultiplet: the full
multiplet
is
\be
({\bf Y}, \l_{\bf Y})
\label{eq:ten}
\ee
where $\l_{\bf Y}$ is the associated Weyl fermion. 
The fermions completing these supermultiplets arise as zero-modes of the 
fermions of five-dimensional heterotic $M$-theory. The action for the
effective,
four-dimensional, $N=1$ theory has been derived in detail in \cite{B32}. 
Here we simply state the result. The relevant terms for a general
discussion of
the superpotential are the kinetic terms for the $S$, $T^I$ and
${\bf Y}$ 
moduli and the bilinear terms of their superpartner fermions. If we 
collectively denote $S$, $T^I$ and ${\bf Y}$ as $Y^{I'}$, where 
$I' = 1,\ldots,h^{1,1}+2$,
and their fermionic superpartners as $\l^{I'}$, then the component
Lagrangian is given by
\bea
\cL_{4D} &=& K_{I'\bar{J}'} \del_u Y^{I'} \del^u 
\bar{Y}^{\bar{J}'} + e^{\k_p^2 K} \left( K^{I'\bar{J}'} 
D_{I'} W \bar{D}_{\bar{J}'} W - 3 \k^2_p | W |^2 \right) \nn \\
& & + K_{I'\bar{J}'} \l^{I'} \delslash \l^{\bar{J}'}
- e^{\k^2_p K / 2} ( D_{I'} D_{J'} W ) \l^{I'} \l^{J'} 
+ \mbox{h.c.}  \label{4Daction}
\eea
Here $\k^2_p$ is the four-dimensional Newton's constant,
\be
K_{I'\bar{J}'} = \del_{I'} \del_{\bar{J}'} K
\ee
are the K\"{a}hler metric and K\"{a}hler potential respectively, and
\be
D_{I'} W = \del_{I'} W + \k^2_p \frac{\del K}{\del Y^{I'}} W
\ee
is the K\"{a}hler covariant derivative acting on the superpotential $W$.
The K\"{a}hler potential, excluding the five-brane translational mode 
${\bf Y}$, was computed in \cite{B32}. This result was extended to
include ${\bf Y}$ in \cite{Derend}. In terms of
the $S$, $T^I$ and ${\bf Y}$ moduli, it is given by
\be
\k^2_p K = - \ln ( S + \bar{S} - \frac{\t}{16} \frac{({\bf Y} + 
\bar{{\bf Y}})^2}{\cT + \bar{\cT}}) 
- \ln \left( \frac{1}{6} 
\sum_{I,J,K=1}^{h^{1,1}}d_{IJK}(T+\bar{T})^I(T+\bar{T})^J(T+\bar{T})^K
\right),
\ee
where $\t$ is the dimensionless parameter
\be
\t = T_5 v_{\CC} (\pi \rho)^2 \k_4^2 .
\ee

It is useful at this point to relate the low energy fields of the
heterotic superstring action derived in Section 4 to the four-dimensional
moduli
derived here from heterotic $M$-theory. Specifically, we note from 
(\ref{eq:two}) that
\be
\hat{g}_{\hat{11}\hat{11}} \mid_{\T = 0} \; = \, \phi^2 \mid_{\T = 0},
\ee
and from (\ref{cymetric}) and (\ref{5dmetric}) that
\be
d s^2_{11} = \cdots + R^2 V^{-2/3}(d y^{\hat{11}})^2.
\ee
Identifying them implies
\be
\phi \mid_{\T = 0} \; = R V^{-1/3}\, . \label{phiisR}
\ee
We will use this identification in the next section.

Following the approach of \cite{BeckerBS} and \cite{HarvMoor}, we will 
calculate the non-perturbative superpotential by computing instanton induced 
fermion bilinear interactions and then comparing these to the fermion bilinear 
terms in the low energy effective supergravity action. In this paper, the 
instanton contribution arises from open supermembranes wrapping on
a product of an interval $I \subset S^1/\Z_2$ and a holomorphic curve 
$\CC \subset CY_3$.
Specifically, we will calculate this instanton contribution to the
two-point
function of the fermions $\l_{\bf Y}$ associated with the ${\bf Y}$
moduli. The two-point function of four-dimensional space-time
fermions $\l_{\bf Y}$ located at positions $y^u_1,y^u_2$ is given by 
the following path integral expression
\be
\langle \l_{\bf Y} (y^u_1) \l_{\bf Y} (y^u_2) \rangle  = 
\int \cD  \Phi  e^{-S_{4D}} \l_{\bf Y} (y^u_1) \l_{\bf Y} (y^u_2)
\cdot \int \cD \hbZ \cD \w e^{-S_{\Si}(\hbZ,\w ; \hbE_{\hbM}^{\; \hbA},
\hbC_{\hbM\hbN\hbP},\BbA_{\BbM},\ubD_{\ubR\ubS} )}, 
\label{pathint}
\ee
where $S_{\Si}$ is the open supermembrane action given in
(\ref{OMaction2}).
Here $\Phi$ denotes all supergravity fields in the $N=1$ supersymmetric 
four-dimensional Lagrangian (\ref{4Daction}) and $\hbZ , \w$ are the 
worldvolume fields on the open supermembrane. In addition,
the path-integral is performed over all supersymmetry preserving 
configurations, $(\hbE_{\hbM}^{\; \hbA},
\hbC_{\hbM\hbN\hbP},\BbA_{\BbM},\ubD_{\ubR\ubS})$, 
of the membrane in the eleven-dimensional \HW
background with a bulk five-brane,
compactified down to four-dimensions on $CY_3 \times S^1/\Z_2$. The integration
will restore $N=1$ four-dimensional supersymmetry.
The result of this calculation is then compared to the terms in
(\ref{4Daction})
proportional to $(D_{{\bf Y}} D_{{\bf Y}} W) \l_{\bf Y} \l_{\bf Y}$
and the non-perturbative contribution to $W$ extracted.

\section{String Action Expansion:}

In this paper, we are interested in the non-perturbative contributions of
open supermembrane instantons to the
two-point function (\ref{pathint}) of chiral fermions in the
four-dimensional effective field theory. In order to preserve $N=1$ 
supersymmetry, the supermembrane must be of the form $\Si = \CC \times 
I$, where curve $\CC \subset CY_3$ is holomorphic and $I \subset S^1/\Z_2$. 
As we have shown in 
previous sections, this is equivalent, in the low energy limit,
to considering the non-perturbative contributions of heterotic superstring 
instantons to the same fermion two-point function in the effective 
four-dimensional theory. Of course, in this setting,
the superstring must wrap completely around a holomorphic curve
$\CC \subset CY_3$ in order for the theory to be $N=1$ supersymmetric.

Since we are interested only in non-perturbative corrections to the
two-point 
function $\langle \l_{\bf Y} (y^u_1) \l_{\bf Y} (y^u_2) \rangle$, the perturbative 
contributions to this function, which arise from the interaction terms in 
the effective four-dimensional action $S_{4D}$ 
in (\ref{pathint}), will not be considered in this paper. Therefore, we
keep
only the kinetic terms of all four-dimensional dynamic fields in $S_{4D}$.
Furthermore, we can perform the functional integrations over all these
fields 
except $\l_{\bf Y}$, obtaining some constant determinant factors which we need
not
evaluate. Therefore, we can rewrite (\ref{pathint}) as
\bea
\langle \l_{\bf Y} (y^u_1) \l_{\bf Y} (y^u_2) \rangle \; &\propto & \,
\int \cD \l_{\bf Y} \, e^{-\int d^4 y 
\l_{\bf Y} \delslash  \l_{\bf Y} } \l_{\bf Y} (y^u_1) \l_{\bf Y} (y^u_2) \nn \\
& & \cdot \int \cD {\Bbb{Z}} \cD \w e^{-S_{\CC}(\BbZ,\w ; \BbE_{\BbM}^{\;
\BbA},
\BbB_{\BbM\BbN},\phi,\BbA_{\BbM},\ubD_{\ubR\ubS})}, \label{pathint2}
\eea
where $S_{\CC}$ is the heterotic superstring action given in
(\ref{HSaction}).
As we will see shortly,
the functional dependence of $S_{\CC}$ on the fields $\l_{\bf Y}$ comes from
the interaction between the superstring fermionic field $\T$
and the five-brane fermion $\cX$  
(from which $\l_{\bf Y}$ is derived in the compactification).
Recall that both of  these fermions are  Weyl spinors in 
ten-dimensions.\footnote{Note 
that in Euclidean space one does not have Majorana-Weyl spinors in 
ten-dimensions.}

Clearly, to perform the computation of the two-point function
(\ref{pathint2}), we must write the action $S_{\CC}$ in terms of its
dynamical 
fields and their interactions with the dimensionally reduced background
fields.
This means that we must first expand all superfield expressions in terms
of
component fields. We will then expand the action in small fluctuations 
around its extrema 
(solutions to the superstring equations of motion), corresponding
to a saddle-point approximation. We will see that because there exists two 
fermionic zero-modes arising from $\T$, their interaction 
with the five-brane fermion $\cX$  will produce a non-vanishing
contribution to
(\ref{pathint2}). Therefore, when performing the path-integrals over
the superstring fields, we must discuss the zero-modes with care.
The next step will be to consider the expression for the
superstring action and to write it in terms of the complex five-brane 
translation modulus.
Finally, we will perform all remaining path integrals in the saddle-point
approximation, obtaining the appropriate determinants. 

We start by expanding the
ten-dimensional superfields in the action $S_{\CC}$ in terms of the 
component fields.

\subsection*{Expanding in Powers of $\T$:}

In this section, for ease of notation, we take the superembedding 
coordinates to be $\BbZ = (X,\T)$ where $(1-\G_{11})\T=0$ as in
(\ref{weylcond}). The required restrictions of $X$ to $\xi$ and $\T$ to
satisfy $(1+i \G_{012345})\T=0$, as in (\ref{addcond}), will
be carried out in the next section along with further gauge fixing
choices.

We begin by rewriting action $S_{\CC}$ in (\ref{HSaction}) as
\be
S_{\CC} = S_S + S_5 + S_{W\!Z\!W},  \label{HetStrAction}
\ee
where
\bea
S_S  (\BbZ ; \BbE_{\BbM}^{\; \BbA}(\BbZ),
\BbB_{\BbM\BbN}(\BbZ),\phi(\BbZ)) &=& T_S
\frac{{Y}}{\pi\rho}\int_{\CC} d^2 \s ( \phi 
\sqrt{\det \del_i \BbZ^{\BbM} \BbE_{\BbM}^{\, A}  \del_j \BbZ^{\BbN}
\BbE_{\BbN}^{\, B} \h_{AB}} \nn \\
& & - \frac{i}{2} \vare^{ij} \del_i \BbZ^{\BbM} \BbE_{\BbM}^{\; \BbA}
\del_j \BbZ^{\BbN} \BbE_{\BbN}^{\; \BbB} \BbB_{\BbB\BbA} ) 
\label{StrAction}
\eea
is the supermembrane bulk action dimensionally reduced on $I\subset S^1/\Z_2$,
\be
S_5 (\BbZ ; \ubD_{\ubR\ubS} (\BbZ))
= \frac{i}{6} T_S \int_{\CC} d^2 \s \vare^{ij} \del_i \BbZ^{\ubR} 
\del_j \BbZ^{\ubS} \ubD_{\ubS\ubR} 
\label{Str5bAction}
\ee
is the action of the boundary string where the membrane meets 
the five-brane and
\bea
S_{W\!Z\!W} (\BbZ,\w ; \BbA_{\BbM} (\BbZ) ) 
&=& - \frac{1}{8 \pi} \int_{\CC} 
 d^2 \s \; \mbox{tr}[\frac{1}{2} \sqrt{g}g^{ij} ( \w_i - \BbA_i )
 \cdot ( \w_j - \BbA_j ) + i \vare^{ij} \w_i \BbA_j ]\nn \\
& & + \frac{1}{24 \pi} \int_{\B} d^3 \hat{\s} i
\hat{\vare}^{\hi\hj\hat{k}}
\Omega_{\hat{k}\hj\hi} (\hat{\w}) . \label{wzwAction}
\eea
is the gauged Wess-Zumino-Witten action on the other boundary string, where
\be 
\BbA_i = \del_i \BbZ^{\BbM} \BbA_{\BbM} (\BbZ).
\ee
Note that this action is a functional of $\BbZ (\s)=(X(\s),\T(\s))$.
We now want to expand the superfields in (\ref{HetStrAction}) in powers of
the fermionic coordinate $\T(\s)$. For the purposes of this paper, we need
only keep terms up to second order in $\T$. We begin with $S_S+S_5$ given
in (\ref{StrAction}) and (\ref{Str5bAction}). 
Using an approach similar to \cite{dewitetal} and
using the results in \cite{Berg}, we find 
that, to the order in $\T$ required, the super-zehnbeins are given by
\be
\BbE_{\BbM}^{\; \BbA} = \left( \ba{cc} E_M^{\, A} 
& \frac{1}{4}\w_M^{\; CD}(\G_{CD})^{\a}_{\, \n}
\T^{\n} \\ - i \G^A_{\; \m \n} \T^{\n} & \d_{\m}^{\a} \ea \right),
\ee
where $E_M^{\, A} (X(\s))$ are the bosonic zehnbeins and $\w_M^{\; CD}(X(\s))$ 
is the ten-dimensional
spin connection, defined in terms of derivatives of $E_M^{\, A}(X)$.
We turn off the gravitino background in this expression for simplicity. We
will discuss below its contribution to the 
two-point function of the fermion related to five-brane translation.  
The super-two-form fields associated with the membrane are, up to the 
order in $\T$ required,
\bea
\BbB_{MN} &=& B_{MN} -\frac{1}{4}\phi\bar{\T}\G_{[M}\G^{CD}\T\w_{N]CD},\nn \\
\BbB_{M\m} &=& -i \phi (\G_M \T)_{\m} , \nn \\
\BbB_{\m \n} &=& 0.
\eea
In addition, we rewrite (\ref{phiisR})
\be
\phi \mid_{\T=0} = R V^{-1/3} . \label{phiisRV}
\ee
Now consider the two-form $\ubD_{\ubR\ubS}$ associated with the five-brane. 
Much of the required information can be obtained from the global supersymmetry 
transformation, which can be read off from the five-brane action 
after choosing the static gauge. The result is
\bea
\ubD_{\ul{r}\ul{s}} &=& \ul{D}_{\ul{r}\ul{s}} - \bar{\cX}\G_{\ul{r}\ul{s}}
\T \nn \\
\ubD_{\ul{r}\ul{\m}} &=& 0 \nn \\
\ubD_{\ul{\m} \ul{\n}} &=& 0,
\eea
where $\cX$ is the 32-component spinor satisfying $(1-\G_{11})\cX=
(1+i\G_{012345})\cX=0$. Its eight independent components form the spinor $\chi$ of
the $(2,0)$ tensor multiplet on the five-brane worldvolume $\ul{M}_6$.
Finally, we find that
\be
\frac{{Y}}{\pi \rho} \phi = \frac{{Y}}{\pi\rho}RV^{-1/3} 
+ \bar{\cX}\T , \label{phiexp}
\ee
where we have used (\ref{phiisRV}). At this point, motivated by the formalism
in \cite{Derend}, we make the field redefinition
\be
\cX = R V^{-1/3} \cX_{{Y}}.
\ee
Expression (\ref{phiexp}) can then be written as
\be
\frac{{Y}}{\pi \rho} \phi = R V^{-1/3} \BbY,
\ee
where
\be
\BbY = \frac{{Y}}{\pi \rho} + \bar{\cX}_{{Y}} \T .
\ee
Hence, fermion $\cX_{{Y}}$ is directly related to the pure
translation modulus ${Y}$.
Substituting these expressions into actions (\ref{StrAction}) and
(\ref{Str5bAction}), they can be written as
\be
S_S + S_5 = S_0 + S_{\T} + S_{\T^2}, \label{StrActionexp}
\ee
where $S_0$ is purely bosonic 
\bea
S_0 (X;E_M^{\;A}(X),\ul{D}_{\ul{r}\ul{s}}(X)) &=& T_S
\frac{{Y}}{\pi\rho}\int_{\CC} d^2 \s ( R V^{-1/3}
\sqrt{\det \del_i X^M \del_j X^N 
E_M^{\; A} E_N^{\; B} \h_{AB}} \nn \\
& & \ \ \ \ \ \ \ \ \ \ \ \ \ \ \ \ \ \ \ \ 
-\frac{i}{2}\vare^{ij}\del_i X^M \del_j X^N B_{NM} ) \nn \\
& & -\frac{i}{6} T_S \int_{\CC} d^2 \s \vare^{ij} \del_i X^{\ul{r}}\del_j 
X^{\ul{s}} \ul{D}_{\ul{s}\ul{r}}  ,
\label{BosAction}
\eea
and $S_{\T}$ and $S_{\T^2}$ are the first two terms (linear and quadratic)
in the $\T$ expansion. Straightforward calculation gives\footnote{In a
space
with Minkowski signature, where the spinors are Majorana-Weyl, the fermion 
product would be $\bar{\cX}_{{Y}}\cV$. However, in Euclidean space, the
fermions are Weyl spinors only and this product becomes the hermitian sum
$\frac{1}{2}(\bar{\cX}_{{Y}} \cV - \bar{\cV}\cX_{{Y}})$.}
\bea
S_{\T} (X,\T;E_M^{\, A}(X),\cX_Y (X)) &=& 
 T_S \int_{\CC} d^2 \s R V^{-1/3} \sqrt{\det \del_i X^M \del_j X^N 
E_M^{\, A} E_N^{\, B} \h_{AB}} \nn \\
& & \cdot \frac{1}{2}(\bar{\cX}_{{Y}}\cV -\bar{\cV}\cX_{{Y}})
\label{vertinAction}
\eea
and
\bea
S_{\T^2} (X,\T; E_M^{\, A}(X)) &=& 
T_S \frac{{Y}}{\pi\rho}\int_{\CC} d^2\s  RV^{-1/3}\sqrt{\det\del_iX^M 
\del_j X^N 
E_M^{\, A} E_N^{\, B} \h_{AB}} \nn \\ & & ( g^{ij}
+ i \e^{ij} ) \bar{\T} \G_i D_j \T , \label{SFqAction}
\eea
where $D_i \T$ is the covariant derivative 
\be
D_i \T = \del_i \T + \del_i X^N \w^{\ A B}_{N}\G_{A B} \T  , 
\label{DiT}
\ee
$\G_i$ is the pullback of the eleven-dimensional Dirac matrices
\be
\G_i = \del_i X^M \G_M , \label{Gpullback}
\ee
and $\cV$ is the vertex operator for the five-brane fermion $\cX_{{Y}}$,
given by
\be
\cV = (1 + \frac{i}{2} \e^{ij}\del_i X^{\ul{r}} \del_j X^{\ul{s}}
\G_{\ul{r}\ul{s}} )\T . \label{vertop}
\ee
The symbol $\e^{ij}$ is the totally antisymmetric tensor in 
two-dimensions, given in terms of the numeric $\vare^{ij}$ by
\be
\e^{ij} = \frac{\vare^{ij}}{\sqrt{\det g_{ij}}} .
\ee
Now consider the expansion of the superfields in $S_{W\!Z\!W}$ given in 
(\ref{wzwAction}). Here, we need only consider the bosonic part of the
expansion
\bea
S_{0W\!Z\!W} (X,\w ; A_M (X),E_M^{\, A}(X) ) 
&=& - \frac{1}{8 \pi} \int_{\CC} 
 d^2 \s \; \mbox{tr}[\frac{1}{2} \sqrt{g}g^{ij} ( \w_i - A_i )
 \cdot ( \w_j - 
A_j ) + i \vare^{ij} \w_i A_j ]\nn \\
& & + \frac{1}{24 \pi} \int_{\B} d^3 \hat{\s} i
\hat{\vare}^{\hi\hj\hat{k}}
\Omega_{\hat{k}\hj\hi} (\hat{\w}) , \label{boswzw}
\eea
where $A_i(\s) = \del_i X^M A_M (X(\s))$ is the bosonic
pullback of $\BbA_{\BbM}$. For example, the expansion of $\BbA_{\BbM}$ to linear
order in $\T$ contains fermions that are not associated with the moduli of 
interest in this paper. Hence, they can be ignored. Similarly, we can show
that all other terms in the $\T$ expansion of $S_{W\!Z\!W}$ are irrelevant to
the problem at hand.

Note that, in terms of the coordinate fields $X$ and $\T$,
the path integral measure in (\ref{pathint2}) becomes\footnote{Since we
are
working in Euclidean space, the spinor fields $\T$ are complex. To be
consistent,
one must use the integration measure $\cD \bar{\T}\cD \T$. 
In this paper, we 
write the integration measure $\cD \T$ as a shorthand for 
$\cD \bar{\T}\cD \T$.}
\be
\cD \BbZ \cD \w = \cD X \cD \T \cD \w . \label{measures}
\ee
We can now rewrite the two-point function as
\bea
\langle \l^I (y^u_1) \l^J (y^u_2) \rangle \; &\propto& \,
\int \cD \l_{\bf Y} \, e^{-\int d^4 y 
\l_{\bf Y} \delslash  \l_{\bf Y}} \l_{\bf Y} (y^u_1) \l_{\bf Y} (y^u_2) \nn \\
& & \cdot \int \cD X \cD \T e^{-(S_0 +S_{\T} +S_{\T^2})} 
\cdot \int \cD \w e^{-S_{0W\!Z\!W}} . \label{pathint2half}
\eea
The last factor
\be
\int \cD \w e^{-S_{0W\!Z\!W}}
\ee
behaves somewhat differently and will be discussed in the next section.  
Here, we simply note that it does not contain the fermion $\l_{\bf Y}$ and,
hence,
only contributes an overall determinant to the superpotential. This
determinant,
although physically important, does not affect the rest of the
calculation,
to which we now turn.
To perform the $X,\T$ path integral, it is essential that we fix
any residual gauge freedom in the $X$ and $\T$ fields.

\subsection*{Fixing the $X$ and $\T$ Gauge:}

As stated at the beginning of the last section, we have, for simplicity,
thus far taken the superembedding coordinates to be $\BbZ=(X,\T)$ where
$(1-\G_{11})\T=0$. Henceforth, however, we must impose the required restrictions
of $X$ to $\xi$ and $\T$ to satisfy $(1+i\G_{012345})\T=0$. In addition, we
will also impose a further choice of gauge. We begin by considering the
bosonic coordinates. As discussed in Section 3, we must take all values of $X^M$
to vanish with the exception of
\be
X^{\ul{r}}(\s) = \xi^{\ul{r}}(\s), \ \ \ \ \ \ \ \ \ \ \ \ \ul{r}=0,1,\ldots,5.
\ee
Having done this, it is convenient to fix the gauge of the non-vanishing 
bosonic coordinates by identifying
\be
X^{\ul{r}'} (\s ) = \d_i^{\ul{r}'} \s^i, \label{bosgaug}
\ee
where $\ul{r}' = 0,1$. This choice, which corresponds to orienting the $X^0$
and $X^1$ coordinates along the string worldvolume, can always be imposed.
This leaves four real bosonic degrees of freedom, which we denote as
\be
X^{u} (\s ) \equiv y^{u}(\s ), \label{bosdof}
\ee
where $u=2,\ldots,5$. Now consider the fermionic coordinate
fields $\T$. First make an two-eight split in the Dirac matrices
\be
\G_{A} = ( \t_{a'} \otimes \tilde{\g} , 1 \otimes \g_{a''} ) , \label{Gamreduc}
\ee
where $a'=0,1$ and $a''=2,\ldots ,9$ are flat indices and $\t_{a'}$ and 
$\g_{a''}$ are the two- and eight-dimensional Dirac matrices, 
respectively. Then $\G_{11}\equiv -i\G_0 \G_1 \cdots \G_9$ can be 
decomposed as
\be
\G_{11} = \tilde{\t} \otimes \tilde{\g}
\ee
where $\tilde{\g}=\g_2 \g_3 \cdots \g_9$ and
\be
\tilde{\t} = -i \t_{0} \t_{1} = \left( \begin{array}{cc} 1 & 0 \\
0 & -1 \end{array} \right) . \label{tildetao}
\ee
More explicitly,
\be
\G_{11} = \left( \begin{array}{cc} 
\tilde{\g} & 0 \\ 0 & - \tilde{\g} \end{array} \right) . \label{gamma11}
\ee
Also note that
\be
-i\G_{012345} = \left( \begin{array}{cc} 
\g_{2345} & 0 \\ 0 & - \g_{2345} \end{array} \right) .  \label{gamma7}
\ee
In general, the Weyl spinor $\T$ can be written in a generic basis as
\be
\T = \left( \begin{array}{c} \T_1 \\ \T_2 \end{array} \right) . \label{GeneralT}
\ee
However, as discussed previously, $\T$ satisfies
\be
\hat{\G}_{\hat{11}} \T = \T \ \ \ \ \ \ \ \ \mbox{and} \ \ \ \ \ \ \ \ \ 
i \G_{012345}\T = \T , \label{gamma7eleven}
\ee
so that the first condition implies
\be
\tilde{\g} \T_1 = \T_1 , \ \ \ \ \ \ \ \ \ \ \ \tilde{\g} \T_2 = - \T_2 ,
\label{T1T2}
\ee
and the second one gives
\be
\g_{2345} \T_1 = \T_1 , \ \ \ \ \ \ \ \ \ \ \ \g_{2345} \T_2 = - \T_2
\ee
From the first equation of (\ref{gamma7eleven}), we conclude that $\T$ is in 
the representation $\mathbf{16}^+$ of $SO(10)$. 
In the presence of the five-brane, $SO(10)$ is 
broken to $SO(4)\times SO(6) \approx SU(2)\times SU(2)\times SO(6)$ under which
\be
\mathbf{16}^+ = (\mathbf{2}^+,\mathbf{1},\mathbf{4}^+) \oplus 
(\mathbf{1},\mathbf{2}^-,\mathbf{4}^-).
\ee
The second projection in (\ref{gamma7eleven}) then implies that $\T$ is in the 
representation $(\mathbf{2}^+,\mathbf{1},\mathbf{4}^+)$. Here, the $\pm$ on 
$\mathbf{2}$ denote $SO(4)$ chirality and the $\pm$ on $\mathbf{4}$
denote $SO(6)$ chirality.
Under the bosonic gauge fixing $X^0 = \s^0$ and $X^1 = \s^1$, $SO(6)$ is
reduced to $SO(4)\times SO(2)$, for which
\be
\mathbf{4}^+ = \mathbf{2}^+ \otimes \mathbf{1}^+ \oplus \mathbf{2}^- 
\otimes \mathbf{1}^- .
\ee
Having applied all the chirality constraints, we can now discuss the 
decomposition of $\T$ under the fermionic gauge fixing conditions.

Recall from our discussion of $\k$-symmetry in Section 2 that, because we can
use the $\k$-invariance of the worldvolume theory to gauge away half of the
independent components of $\T$, only half of these components
represent physical degrees of freedom. For the superstring in Euclidean space,
we can define the projection operators
\be
P_{\pm} = \frac{1}{2} (1 \pm \frac{i}{2\sqrt{g}}\vare^{ij}\Pi_i^{\, A}
\Pi_j^{\, B} \G_{AB})
\ee
and write
\be
\T = P_+ \T + P_- \T .
\ee
Now, note from (\ref{kappa}) that $P_+ \T$ can be gauged away while
the physical degrees of freedom are given by $P_- \T$. Using (\ref{tildetao}), 
it follows that $\T_2$ in (\ref{GeneralT}) can
be gauged to zero, leaving only $\T_1$ as the physical degrees of freedom. 
We thus can fix the fermion gauge so that
\be
\T = \left( \begin{array}{c} \q \\ 0 \end{array} \right), \label{Theta}
\ee
where $\q$ satisfies
\be
\tilde{\g} \q = \q , \ \ \ \ \ \ \ \ \ \ \ \ \ \ \
\g_{2345} \q = \q , \label{thetaugammas}
\ee
and transforms in the representation
\be
(\mathbf{2}^+,\mathbf{1},\mathbf{2}^+,\mathbf{1}^+) 
\ee
under $SU(2)\times SU(2)\times SO(4) \times SO(2)$. This corresponds to 
choosing $\mathbf{1}^+$ under the $SO(2)$ chirality of the string
worldsheet, which implies that the physical $\T$ is the right-moving mode.
We conclude that the physical
degrees of freedom contained in $\BbZ = (X,\T)$ are
\be
y^u (\s), \ \ \ \ \ \ \ \ \ \ \ \ \ \ \  \q^A_{\a} (\s),
\ee
where $u=2,\ldots,5$ indexes $R_4$, $A=1,2$ is the $SU(2)$ index and $\a$ 
denotes the $\mathbf{2}^+$ of the $SO(4)$ symmetry of $R_4$.
Therefore, the $X,\T$ path-integral measures in (\ref{pathint2half}) 
must be rewritten as
\be
\cD X \cD \T \propto \cD y \cD \q ,
\ee
where there is an unimportant constant of porportionality representing the 
original gauge redundancy.\footnote{Here, again, we write $\cD \q$ as a 
shorthand for $\cD \bar{\q} \cD \q$.}

\subsection*{Equations of Motion:}

We can now rewrite the two-point function (\ref{pathint2half}) as
\bea
\langle \l_{\bf Y} (y^u_1) \l_{\bf Y} (y^u_2) \rangle \; &\propto & 
\int \cD \l_{\bf Y} \, e^{-\int d^4 y 
\l_{\bf Y} \delslash  \l_{\bf Y}} \l_{\bf Y} (y^u_1) \l_{\bf Y} (y^u_2) \nn \\
& & \cdot \int \cD y \cD \q e^{-(S_0 +S_{\T} +S_{\T^2})} 
\cdot \int \cD \w e^{-S_{0W\!Z\!W}} . \label{pathint3}
\eea
In this paper, we want to use a saddle-point approximation to evaluate
these
path-integrals. We will consider small fluctuations $\d y$ and $\d\q$
of the superstring degrees of freedom around a solution $y_0$ and 
$\q_0$ to the equations of motion
\be
y = y_0 + \d y , \ \ \ \ \ \ \ \ \ \ 
\q = \q_0 + \d \q . \label{allfluc} 
\ee
However, before expanding the action using (\ref{allfluc}), we
need to discuss the equations of motion for the fields $y$ and $\q$, 
as well as their zero-modes.

Consider first the equations of motion for the bosonic fields $y(\s)$. The
bosonic action (\ref{BosAction}) can be written as
\be
S_0 = T_S \frac{{Y}}{\pi\rho}\int_{\CC} d^2 \s ( R V^{-1/3}
\sqrt{\det g_{ij}} + \frac{i}{2} \vare^{ij} b_{ij} )
- T_S \int_{\CC} d^2 \s\frac{i}{6} \vare^{ij} \ul{d}_{ij}   , 
\label{BosAct}
\ee
where
\be
g_{ij} = \del_i X^{\ul{r}} \del_j X^{\ul{s}} g_{\ul{r}\ul{s}} , \ \ \
 \ \ \ \ \ \ \ b_{ij} = \del_i X^{\ul{r}} \del_j X^{\ul{s}} B_{\ul{r}\ul{s}},
 \ \ \ \ \ \ \ \ \ \ \ul{d}_{ij} = \del_i X^{\ul{r}} \del_j X^{\ul{s}} 
\ul{D}_{\ul{r}\ul{s}} . \label{gandb}
\ee
We now assume that the background two-form field $B_{MN}(X)$ satisfies $dB=0$.
This can be done if we neglect corrections of order $\a'$.
Then, locally, $B = d \Lambda$, where $\Lambda$ is a one-form.
Thus, the second term in (\ref{BosAct}) can be written as a total derivative
and so does not contribute to the equations of motion. 
Next, note that, similarly, the five-brane two-form 
$\ul{D}_{\ul{r}\ul{s}}(X)$ satisfies $d\ul{D}=0$. This can be seen as follows.
Recall from (\ref{2.21})
that $(d\ul{D})_{\ul{r}\ul{s}\ul{t}}=\ubC_{\ul{r}\ul{s}\ul{t}}$. 
However, the field components $\BbC_{MNP}$ vanish in the low energy
limit of heterotic $M$-theory because of their $\Z_2$ properties. The result
then follows. Therefore, locally, $\ul{D} = d \ul{\L}$, where $\ul{\L}$ is a one-form.
Hence, the third term in (\ref{BosAct}) can also be written as a total
derivative
and so does not contribute to the equations of motion. Varying the action,
we obtain the bosonic equations of motion
\bea
\frac{1}{2} \sqrt{\det g_{ij}} g^{kl} \del_k X^{\ul{r}} \del_l X^{\ul{s}} 
\frac{\del g_{\ul{r}\ul{s}}}{\del X^{\ul{t}}}
& & \nn \\
- \del_k (\sqrt{\det g_{ij}} g^{kl} \del_l X^{\ul{r}} g_{\ul{r}\ul{t}}) &=& 0,
\label{stringeom}
\eea
where $g^{ij}$ is the inverse of the induced metric $g_{ij}$, 
$g^{ij}g_{jk}=\d^i_k$. Now fix the bosonic gauge (\ref{bosgaug}) and
choose a system of coordinates such that the
metric tensor restricted to the holomorphic curve $\CC$ can be written locally as
\be
g_{\ul{r}\ul{s}} \mid_{\CC} = \left( \ba{cc} h_{\ul{r}'\ul{s}'} (\s) & 0 \\ 
0 & \h_{uv} \ea \right), \label{gsplit}
\ee
where $\h_{uv}$ is the flat metric of $R_4$.
Then equation (\ref{stringeom}) becomes 
\be
\del_k \left( \sqrt{\det g_{ij}}  \d_{\ul{r}'}^k\d_{\ul{s}'}^{l} 
h^{\ul{r}'\ul{s}'} \del_l y_0^u  \right) = 0. \label{eomforX}
\ee

Next, consider the equations of motion for the fermionic degrees of
freedom. In action (\ref{StrAction}) the terms that contain $\T$ are 
(\ref{vertinAction}) and (\ref{SFqAction}), whose sum can be written as
\be
2 T_S \frac{{Y}}{\pi\rho}\int_{\CC} d^2 \s R V^{-1/3} \sqrt{\det
g_{ij}} \bar{\T} \G^i D_i \T +\frac{1}{2} T_S \int_{\CC} d^2 \s R V^{-1/3}
\sqrt{\det g_{ij}} (\bar{\cX}_{{Y}}\cV-\bar{\cV}\cX_{{Y}}),
\label{sumfact}
\ee
where we have fixed the gauge as in (\ref{bosgaug}) and (\ref{Theta}), 
so that $\cV$ is given by 
\bea
\cV &=& (1 + \frac{i}{2} \e^{ij}\del_i X^{\ul{r}} \del_j X^{\ul{s}}
\G_{\ul{r}\ul{s}} )\T  \nn \\
&=& 2 \T \label{vertposfix}
\eea 
It follows from the gauge fixing condition (\ref{Theta})
that only half of the eight independent components of the five-brane fermion 
$\cX_{{Y}}$ couple to the physical degrees of freedom in $\T$, namely 
\be
P_+\cX_{{Y}}=\frac{1}{2} 
(1 + i \G_{01}) \cX_{{Y}}\equiv \cX_{{Y}}^+ .
\ee
In fact, the equations of motion for $\T$ are given by
\be
2 \frac{{Y}}{\pi\rho}  \G^i D_{0i} \T_0 = \cX_{{Y}}^+, 
\label{eomforTheta}
\ee
where we have used (\ref{Gamreduc}) and
\be
D_{0i} \T_0 = \del_i \T_0 + \d_i^{\ul{r}'}  
\w^{\ \breve{K} \breve{L}}_{\ul{r}'} \G_{\breve{K}\breve{L}} \T_0 .
\ee
Of course, we must consider only the physical degrees of freedom $\q_0$ in $\T_0$.

\subsection*{Zero-Modes:}

The saddle-point calculation of the path-integrals
$\cD y$ and $\cD \q$ around a solution to the equations of motion can 
be complicated by the occurrence of zero-modes.
First consider bosonic solutions $y_0^{u}(\s)$, $u=2,\ldots,5$ of the
equations of motion (\ref{eomforX}). By construction, all such solutions 
are maps from a holomorphic curve $\CC$ to $R^4$.  
 Clearly, these can take any value in $R_4$, so we can
write
\be
y_0^u \equiv x^u , \label{boszeromodes}
\ee
where $x^u$ are coordinates of $R_4$. Therefore, any solution $y_0^{u}(\s)$
of the equations of motion will always have these four translational
zero-modes. Are additional zero-modes possible? To avoid this possibility,
we will assume in this paper that
\be
\CC = \BbC \BbP^1 = S^2 ,
\ee
where $S^2$ are rigid spheres isolated in $CY_3$.
It follows that for a saddle-point calculation of the
path-integrals around a rigid, isolated sphere, the bosonic measure can be
written as
\be
\cD y^u = d^4 x \, \cD \d y^u ,
\ee
where we have expanded
\be
y^{u} = y^{u}_0 + \d y^{u}  \label{bosfluc}
\ee
for small fluctuations $\d y^{u}$.

Now consider fermionic solutions $\q_0$ of the equation of motion 
(\ref{eomforTheta}). To any $\T_0$ can always be added a solution of the
homogeneous six-dimensional Dirac equation
\be
\Gamma^i D_{0i} \T' = 0 . \label{diraceq}
\ee
This equation has the general solution 
\be
\T' = \vart \otimes \h_- , \label{zeromodeansatz}
\ee
where $\h_-$ is the covariantly constant spinor on $CY_3$, which is broken
by the embedding of the membrane as discussed in \cite{LOPR}, restricted to $\CC$ 
and $\vart$ is an arbitrary Weyl spinor satisfying the Weyl equation in $R_4$. 
Note that $\vart$ has negative four-dimensional chirality, since $\T'$ satisfies
$(1-\G_{11})\T'=0$.
Therefore, any solution $\q_0$ of the
equations of motion will always have two complex component fermion
zero-modes $\vart^{\a}$, $\a=1,2$. The rigid, isolated sphere has no
additional
fermion zero-modes. Hence, for a saddle-point calculation of the path
integrals
around a rigid, isolated sphere the fermionic measure can be written as
\be
\cD \q = d \vart^1 d \vart^2 \, \cD \d \q ,
\ee
where we have expanded
\be
\q = \q_0 + \d \q
\ee
for small fluctuations $\d \q$. To conclude, in the saddle-point
approximation
the $y,\q$ part of the path integral measure can be written as
\be
\cD y^u \cD\q =d^4 x\, d \vart^1 d \vart^2 \,\cD\d y^u
\cD\d\q .
\ee

\subsection*{Saddle-Point Calculation:}

We are now ready to calculate the two-point function (\ref{pathint3}),
which can be rewritten as
\bea
\langle \l_{\bf Y} (y^u_1) \l_{\bf Y} (y^u_2) \rangle \; &\propto &
\int \cD \l_{\bf Y} \, e^{-\int d^4 y 
\l_{\bf Y} \delslash  \l_{\bf Y}} \l_{\bf Y} (y^u_1) \l_{\bf Y} (y^u_2) 
\nn \\
& & \cdot \int d^4 x\, d \vart^1 d \vart^2 \,\cD\d y^u
\cD\d\q \, e^{-(S_0 + S_{\T} + S_{\T^2})} \nn \\ & & \cdot 
\int \cD \w \,  e^{-S_{0W\!Z\!W}}. \label{pathint4}
\eea
Substituting the fluctuations (\ref{allfluc}) around
the solutions $y_0$ and $\q_0$ into
\be
\S =  S_0 + S_{\T} + S_{\T^2} ,
\ee
we obtain the expansion
\be
\S = \S_0 + \S_2  ,
\ee
where, schematically
\be
\S_0 = \S \mid_{y_0 , \q_0} 
\ee
and
\be
\S_2 = \frac{\d^2 \S}{\d y \d y} \mid_{y_0 , \q_0} (\d y)^2
+ 2 \frac{\d^2 \S}{\d y\d\q} \mid_{y_0 , \q_0} (\d y\d\q)
+ \frac{\d^2 \S}{\d\q\d\q} \mid_{y_0 , \q_0} (\d\q)^2 .
\ee
The terms in the expansion linear in $\d y$ and $\d\q$ each vanish by the
equations of motion. To avoid further complicating our notation, we state
in
advance the following simplifying facts. First, note that all terms in
$\S_2$
contribute to the two-point function to order $\a'$ on the superstring
worldsheet. Therefore, we should evaluate these terms only to classical
order
in $y^u_0$ and $\q_0$. To classical order, one can take $\q_0 = 0$
since,
to this order, the background $\cX_{{Y}}^-$ field on the right-hand side of 
(\ref{eomforTheta}) vanishes. Therefore, $\S_2$ simplifies to
\be
\S_2 = \frac{\d^2 \S}{\d y \d y} \mid_{y_0 , \q_0 = 0} (\d y)^2
+ \frac{\d^2 \S}{\d\q\d\q} \mid_{y_0 , \q_0 = 0} (\d\q)^2 .
\ee
It is useful to further denote
\be
\S_0 = \S_0^y + \S_0^{\q} , 
\ee
where
\be
\S_0^y = ( S_0 ) \mid_{y_0}, \ \ \ \ \ \ \ \ \ \ \ \ \
\S_0^{\q} = ( S_{\T} + S_{\T^2} ) \mid_{y_0,\q_0} , \label{sysq}
\ee
and to write
\be
\S_2 = \S_2^y + \S_2^{\q},
\ee
with
\be
\S_2^y = \frac{\d^2 \S}{\d y \d y} \mid_{y_0 , \q_0 =0} (\d y)^2 ,
\ \ \ \ \ \ \ \ \ \ \
\S_2^{\q} = \frac{\d^2 \S}{\d\q\d\q} \mid_{y_0 , \q_0 =0} (\d\q)^2 .
\label{s2}
\ee
We can then rewrite two-point function (\ref{pathint4}) as
\bea
\langle \l_{\bf Y} (y^u_1) \l_{\bf Y} (y^u_2) \rangle \; &\propto &
\int \cD \l_{\bf Y} \, e^{-\int d^4 y 
\l_{\bf Y} \delslash  \l_{\bf Y}} \l_{\bf Y} (y^u_1) \l_{\bf Y} (y^u_2) \nn \\
& & \cdot \int d^4 x\, e^{-\S_0^y} \cdot \int d \vart^1 d \vart^2 \, 
e^{-\S_0^{\q}} \nn \\ 
& & \cdot \int \cD\d y^u \, e^{-\S_2^y} \cdot
\int \cD\d\q \, e^{-\S_2^{\q}} \cdot 
\int \cD \w \,  e^{-S_{0W\!Z\!W}}. \label{pathint5}
\eea
We will now evaluate each of the path-integral factors in this expression
one
by one. We begin with $\int d^4 x \, e^{-\S_0^y}$.

\subsection*{The $\S_0^y$ Term:}

It follows from (\ref{sysq}) that $\S_0^y$ is simply $S_0$, given in
(\ref{BosAct}) and (\ref{gandb}), evaluated at a solution of the
equations of motion $y_0^{u}$. That is
\be
\S_0^y = T_S \frac{{Y}}{\pi\rho}\int_{\CC} d^2 \s ( R V^{-1/3}
\sqrt{\det g_{ij}} + \frac{i}{2} \vare^{ij} b_{ij} )
+\frac{i}{6} T_S \int_{\CC} d^2 \s 
\vare^{ij} \ul{d}_{ij} , \label{threechecks}
\ee
where
\be
g_{ij} = \del_i y_0^{\ul{r}} \del_j y_0^{\ul{s}} g_{\ul{r}\ul{s}} , \ \ \
 \ \ \ \ \ \ \ b_{ij} = \del_i y_0^{\ul{r}} \del_j y_0^{\ul{s}} B_{\ul{r}\ul{s}},
 \ \ \ \ \ \ \ \ \ \ \ul{d}_{ij} = \del_i y_0^{\ul{r}} \del_j y_0^{\ul{s}} 
\ul{D}_{\ul{r}\ul{s}} .
\label{gandb2}
\ee
Let us evaluate the term involving $g_{ij}$. To begin, we note that
\be
\int_{\CC} d^2 \s \sqrt{\det g_{ij}} = \frac{1}{2} \int_{\CC} d^2 \s
\sqrt{g}
g^{ij} \del_i y_0^{\ul{r}} \del_j y_0^{\ul{s}} g_{\ul{r}\ul{s}} , 
\label{onecheck}
\ee
where the first term is obtained from the second using the worldvolume
metric
equation of motion. Noting that $g_{ij}$ is conformally flat and going to
complex coordinates $z=\s^0 + i \s^1$, $\bar{z}=\s^0 - i\s^1$, it follows
from (\ref{onecheck}) that
\be
\int_{\CC} d^2 \s \sqrt{\det g_{ij}} = \frac{1}{2} \int_{\CC} d^2 z
\del_z y_0^{\ul{r}} \del_{\bar{z}} y_0^{\ul{s}} \w_{\ul{r}\ul{s}}
=\frac{1}{2} \int_{\CC} d^2 z \omega_{z\bar{z}} , 
\label{twochecks}
\ee
where $\w_{\ul{r}\ul{s}}=ig_{\ul{r}\ul{s}}$ is the K\"{a}hler form 
restricted to $\CC$. Using the expansion (\ref{expan}) and the orthonormal
conditions (\ref{CCandw}), (\ref{w'isortho}), it follows 
from (\ref{twochecks}) that
\be
\int_{\CC} d^2 \s RV^{-1/3} \sqrt{\det g_{ij}} = \frac{v_{\CC}}{2}
\mbox{Re} \cT . \label{g1}
\ee
Next consider the second term in (\ref{threechecks}) involving $b_{ij}$.
Note that
\be
\frac{i}{2} \int_{\CC} d^2 \s \vare^{ij} b_{ij} = 
\frac{i}{2} \int_{\CC} d^2 z
\del_z y_0^{\ul{r}} \del_{\bar{z}} y_0^{\ul{s}} B_{\ul{r}\ul{s}}
= \frac{i}{2} \int_{\CC} d^2 z B_{z\bar{z}} . \label{porra2}
\ee
Recall from (\ref{Bisimtao}) that
\be
B_{z\bar{z}} = \mbox{Im} \cT \w_{\CC z\bar{z}} + \cdots ,
\ee
where the dots indicate terms that vanish upon integration over
$\CC$. It follows from (\ref{CCandw}) and (\ref{porra2}) that
\be
\frac{i}{2} \int_{\CC} d^2 \s  
\vare^{ij} b_{ij} = \frac{i}{2} v_{\CC} \mbox{Im} \cT .
\label{b2}
\ee
Finally, consider the third term in (\ref{threechecks})
involving $\ul{d}_{ij}$. First, we note that
\be
\frac{i}{6} \int_{\CC} d^2 \s \vare^{ij} \ul{d}_{ij} = 
\frac{i}{6} \int_{\CC} d^2 z
\del_z y_0^{\ul{r}} \del_{\bar{z}} y_0^{\ul{s}} \ul{D}_{\ul{r}\ul{s}}
= \frac{i}{6} \int_{\CC} d^2 z \ul{D}_{z\bar{z}} . \label{porra}
\ee
Remembering from (\ref{disa}) that
\be
\ul{D}_{z\bar{z}} = 3 a \w_{\CC z\bar{z}} ,
\ee
it follows from (\ref{CCandw}) and (\ref{porra}) that
\be
\frac{i}{6} \int_{\CC} d^2 \s  \vare^{ij} \ul{d}_{ij} = 
\frac{i}{2} v_{\CC} a . \label{d3}
\ee
Putting (\ref{g1}), (\ref{b2}) and (\ref{d3}) together in (\ref{threechecks}), 
we see that
\be
\S_0^y = \frac{T}{2} \left( \frac{{Y}}{\pi\rho}\mbox{Re}\cT 
+ i (a + \frac{{Y}}{\pi\rho}\mbox{Im}\cT) \right) ,
\ee
where
\be
T = T_S v_{\CC} = T_M \pi \rho \, v_{\CC} 
\ee
is a dimensionless parameter. Recalling from (\ref{bigY}) that the 
${\bf Y}$ modulus is defined by
\be
{\bf Y} = \frac{{Y}}{\pi\rho} \mbox{Re} \cT + i(a+
\frac{{Y}}{\pi\rho} \mbox{Im} \cT ) , \label{thisisY}
\ee
it follows that we can write $\S_0^y$ as
\be
\S_0^y = \frac{T}{2} {\bf Y}.
\ee
We conclude that the $\int d^4 x e^{-\S_0^y}$ factor in the path-integral
is given by
\be
\int d^4 x \, e^{-\S_0^y} = \int d^4 x e^{-\frac{T}{2}
{\bf Y} } .
\ee
We next evaluate the path integral factor $\int d \vart^1 d \vart^2 
e^{-\S_0^{\q}}$.

\subsection*{The $\S_0^{\q}$ Term and the Fermionic Zero-Mode Integral:}

It follows from (\ref{sysq}) that $\S_0^{\q}$ is the sum of $S_{\T}$ and
$\S_{\T^2}$, given in (\ref{sumfact}), evaluated at a solution of the
equations of motion $y_0^u$, $\q_0$. Varying (\ref{sumfact}) with
respect to $\bar{\T}$ leads to the equation of motion (\ref{eomforTheta}).
Inserting the equation of motion into (\ref{sumfact}), we find
\be
\S_0^{\q} = T_S \int_{\CC} d^2 \s R V^{-1/3} \sqrt{\det g_{ij}}
\bar{\cX}_{{Y}}\T_0 .  \label{s0q}
\ee
As discussed above, any solution $\T_0$ can be written as the sum
\be
\T_0 = \hat{\T}_0 + \T' ,
\ee
where $\T'$ is a solution of the purely homogeneous Dirac equation 
(\ref{diraceq}) and has the form (\ref{zeromodeansatz}). Since, in the
path-integral, we must integrate over the two zero-modes $\vart^{\a}$,
$\a=1,2$ in $\T'$, it follows that terms involving $\hat{\T}_0$ can never
contribute to the fermion two-point function. Therefore, when computing
the superpotential, one can simply drop $\hat{\T}_0$. Hence, $\S_0^{\q}$
is given by (\ref{s0q}) where $\T_0$ is replaced by $\T'$.

Next, we note that the Kaluza-Klein Ansatz for the ten-dimensional  
fermion $\cX_{{Y}}$ is given by
\be
\cX_{{Y}} = - i \l_{{Y}} \otimes \h_- , \label{cygravit}
\ee
where $\l_{{Y}}(y^u)$ are the fermionic superpartners of the 
complex modulus ${Y}$ with four-dimensional negative chirality.
Using (\ref{zeromodeansatz}) and (\ref{cygravit}), one can evaluate the product
$\bar{\cX}_{{Y}}\T'$, which is found to be
\be
\bar{\cX}_{{Y}}\T' = -i \cdot (\l_{{Y}} \vart),
\ee
where $\l_{{Y}}\vart = \l_{{Y}\a}\vart^{\a}$ and we used the fact 
that the $CY_3$ covariantly constant spinor $\h_-$ is normalized to one. Substituting 
this expression into (\ref{s0q}) and using (\ref{g1}) then gives
\be
\S_0^{\q} = T \, \mbox{Re} \cT \l_{{Y}} \vart . \label{fbcont}
\ee
However, we are note quite finished. Thus far, in this section, we have ignored
the gravitino for notational simplicity and because we have presented the 
gravitino formalism in detail in \cite{LOPR}. Using that formalism, it is
straightforward to compute the contribution of the gravitino to $\S_0^{\q}$,
which we find to be
\be
T \frac{{Y}}{\pi\rho} \l_{\cT} \vart, \label{gravcont}
\ee
where $\l_{\cT}$ is the fermionic superpartner of modulus $\cT$ discussed in
Section 5. Combining (\ref{fbcont}) with (\ref{gravcont}), we have the
complete result that
\be
\S_0^{\q} = T \l_{{\bf Y}} \vart ,
\ee
where
\be
\l_{{\bf Y}} = \mbox{Re} \cT \l_{{Y}} + {Y} \l_{\cT} 
\ee
is the fermionic superpartner of modulus ${\bf Y}$. It is gratifying that
this expression for $\l_{{\bf Y}}$, as well as expression (\ref{thisisY}) 
for ${\bf Y}$, are consistent with those found, in a different context,
in \cite{Derend}.
It follows that the $\int d \vart^1 d \vart^2 \, e^{-\S_0^{\q}}$ factor in
the path-integral is
\be
\int d \vart^1 d \vart^2 e^{- \S_0^{\q}} = \int d \vart^1 d \vart^2 
e^{ - T  \, \l_{\bf Y} \vart } .
\ee
Expanding the exponential, and using the properties of the Berezin
integrals, we
find that
\be
\int d \vart^1 d \vart^2 e^{- \S_0^{\q}} = \frac{T^2}{2}
\, \l_{\bf Y} \l_{\bf Y} ,
\ee
where we have suppressed the spinor indices on $\l_{\bf Y} \l_{\bf Y}$.
Collecting the results we have obtained thus far, two-point function 
(\ref{pathint5}) can now be written as
\bea
\langle \l_{\bf Y} (y^u_1) \l_{\bf Y} (y^u_2) \rangle \; &\propto &
\int \cD \l_{\bf Y} \, e^{-\int d^4 y 
\l_{\bf Y} \delslash  \l_{\bf Y}} \l_{\bf Y} (y^u_1) \l_{\bf Y} (y^u_2) \nn \\ 
& & \cdot \int d^4 x\, e^{- \frac{T}{2}{\bf Y} (x)}
\, \l_{\bf Y} (x) \l_{\bf Y} (x) \nn \\ 
& & \cdot \int \cD\d y^u \, e^{-\S_2^y} \cdot
\int \cD\d\q \, e^{-\S_2^{\q}} \cdot 
\int \cD \w \,  e^{-S_{0W\!Z\!W}}. \label{pathint6}
\eea
Next, we evaluate the bosonic path-integral factor $\int \cD \d y^u
e^{- \S_2^y}$.

\subsection*{The $\S_2^y$ Quadratic Term:}

It follows from (\ref{s2}) that $\S_2^y$ is simply the quadratic term in
the $y=y_0 + \d y$ expansion of $\S_0$, given in (\ref{BosAct}) and 
(\ref{gandb}). Note that $S_{\T} + S_{\T^2}$ does not contribute since the 
second derivative is to be evaluated for $\q_0 = 0$. Furthermore, since this
contribution to the path-integral is already at order $\a'$, $\S_0$ should
be evaluated to lowest order in $\a'$. As discussed above, to lowest order
$dB=0$ and, hence, the $b_{ij}$ term in (\ref{BosAct}) is a total
divergence which can be ignored. In addition, as discussed above, 
$d\ul{D}=0$ and, thus, the $\ul{d}_{ij}$ term in (\ref{BosAct}) 
is also a total divergence which can be ignored.
Performing the expansion in what is left, we find that
\be
S_2^y = T_S \frac{{Y}}{\pi\rho} \int_{\CC} d^2 \s R V^{-1/3} 
\sqrt{\det g_{ij}} \left( \frac{1}{2} 
g^{ij} (D_i \d y^{u} )(D_j \d y^{v})\h_{uv} \right). \label{s2y}
\ee
The induced covariant derivative of $\d y$ is a simple ordinary derivative  
\be
D_i \d y^{u} = \del_i \d y^{u} + \w_{i \; \, v}^{\, u} \d y^{v}=\del_i \d
y^{u},
\label{indcov}
\ee
since the connection components vanish along $R^4$. 
Integrating the derivatives by parts then gives
\bea
S_2^y &=& T_S \frac{{Y}}{\pi\rho}\int_{\CC} d^2 \s  R V^{-1/3}
\left( - \frac{1}{2} \d y^u [ \h_{uv} \sqrt{g} g^{ij} \cD_i \del_j ] 
\d y^v \right)
\eea
where the symbol $\cD_i$ indicates the covariant derivative with respect
to the worldvolume connection on
$\CC$. Generically, the fields $R,V$ and ${Y}$ are functions of $x^u$.
However,
as discussed above, at the level of the quadratic contributions to the
path-integrals all terms should be evaluated at the classical values of
the background fields. Since $R,V$ and ${Y}$ are moduli, these classical
values can be taken to be constants, rendering ${Y}RV^{-1/3}$ independent
of $x^u$. Hence, the factor $T_S \frac{{Y}}{\pi\rho}RV^{-1/3}$ can 
simply be absorbed by a redefinition of
the $\d y$'s. Using the relation
\be
\int\cD\d y\;e^{-\frac{1}{2}\int d^2\s\,\d y\cO\d y} \propto
\frac{1}{\sqrt{\det \cO}} ,
\ee
we conclude that
\be
\int \cD \d y^u e^{- \S_2^y} \propto \frac{1}{\sqrt{\det \cO_1}}
\ee
where
\bea
\cO_1 &=& \h_{uv} \sqrt{g} g^{ij} \cD_i \del_j 
\label{Os}
\eea
We next turn to the evaluation of the $\int \cD \d \q \,e^{- \S_2^{\q}}$
factor
in the path-integral.

\subsection*{The $\S_2^{\q}$ Quadratic Term:}

It follows from (\ref{s2}) that $\S_2^{\q}$ is the quadratic term in the
$\q = \q_0 + \d \q$ expansion of $S_{\T^2}$, given in (\ref{SFqAction}).
Note that $S_0 + S_{\T}$ does not contribute. Performing the expansion and
taking into account the gauge fixing condition, we find that
\be
\S_2^{\q} = 2T_S\frac{{Y}}{\pi\rho} \int_{\CC} d^2 \s R V^{-1/3} \sqrt{\det g_{ij}}
\d \bar{\T} \G^i D_i \d \T . \label{quadtheta}
\ee
One must now evaluate the product $\d\bar{\T} \G^i D_i \d\T$ in terms of
the gauged-fixed quantities $\d\q$. We start by rewriting
\bea
\d \bar{\T} \G^i D_i \d \T &=& g^{ij} \del_j X^{\ul{r}} \d \bar{\T} \G_{\ul{r}}
\del_i \d \T \nn \\
& & + g^{ij} \del_j X^{\ul{r}} \del_i X^{\ul{s}} \w^{\ A B}_{\ul{s}} 
\d \bar{\T} \G_{\ul{r}} \G_{A B} \d \T , \label{Thet21}
\eea
where $A=(a',a'')$ and we have used the restrictions on fields $X^M (\s)$.
After fixing the gauge freedom of the bosonic fields $X^{\ul{r}} (\s)$
as in (\ref{bosgaug}), expression (\ref{Thet21}) becomes
\bea
\d \bar{\T} \G^i D_i \d \T &=& g^{ij} \d_j^{m'} e_{m'}^{\ a'} \d \bar{\T} 
\G_{a'} \del_i \d \T + g^{ij} e_u^{\ k} \del_j y^u \d \bar{\T} 
\G_k \del_i \d \T \nn \\
& & + g^{ij} \d_j^{m'} e_{m'}^{\ a'} \d_i^{n'} \w^{\ A B}_{n'}
\d \bar{\T} \G_{a'} \G_{A B} \d \T \nn \\
& & + g^{ij} e_u^{\ k} \del_j y^u \d_i^{m'} \w^{\ A B}_{m'} 
\d \bar{\T} \G_k \G_{A B} \d \T ,
\label{Thet22}
\eea
where $k=2,3,4,5$ are flat indices in $R_4$. 
We see that we must evaluate the fermionic products
\be
\d \bar{\T} \G_{a'} \del_i \d \T, \ \ \ \ \d \bar{\T} \G_k \del_i \d \T, 
\ \ \ \ \d \bar{\T} \G_{a'} \G_{A B} \d \T, \ \ \ \ \d \bar{\T} \G_k 
\G_{A B} \d \T 
\ee
in terms of $\d \q$. After fixing the fermionic gauge according to 
(\ref{Theta}), we can compute the relevant terms in the expression 
(\ref{Thet22}). Consider a product of the type 
$\d \bar{\T}M\d\T$, where $M$ is a $32\times 32$ matrix-operator,
\be
M = \left( \ba{cc} M_1 & M_2 \\ M_3 & M_4 \ea \right) .
\ee
Using (\ref{Gamreduc}) and (\ref{Theta}), we have
\be
\d \bar{\T} M \d \T = \d \T^{\dagger} M \d \T =  \d \q^{\dagger} M_1 \d \q
\ee
Therefore, using (\ref{Gamreduc}), we have the following results
\be
\ba{rl} \d \bar{\T}\G_{a'}\del_i\d\T= 0,
 & \d \bar{\T}\G_{a''}\del_i\d\T=\d\q^{\dagger}\g_{a''}\del_i\d\q , \\
\d\bar{\T}\G_{a'}\G_{b' c'}\d\T=0, & 
\d \bar{\T}\G_{a'}\G_{b' a''}\d\T=(\d_{a' b'}-i\vare_{a' b'})\d\q^{\dagger}
\g_{a''}\d\q , \\
\d\bar{\T}\G_{a'}\G_{a'' b''}\d\T=0, &
 \bar{\T}\G_{a''}\G_{a' b'}\T=-i\vare_{a' b'}\d\q^{\dagger}\g_{a''}\d\q , \\
\d\bar{\T}\G_{a''}\G_{a' b''}\d\T=0, & 
\d \bar{\T}\G_{a''}\G_{b'' c''}\d\T=\d\q^{\dagger}\g_{a''}\g_{b'' c''}\d\q ,\ea
\ee
with $a''=(k,K)$, $k=2,3,4,5$ are flat indices in $R_4$ and $K=6,7,8,9$ are 
flat indices in the supspace $CY_{\perp} \subset CY_3$ orthogonal to
$\CC$. Substituting these expressions into (\ref{Thet22}) yields
\bea
\d\bar{\T}\G^i D_i\d\T &=&  
g^{ij} e_u^{\ k} \del_j y^u [ \d\q^{\dagger}\g_k\del_i\d\q
-i \d_i^{m'}\w^{\ a' b'}_{m'} \vare_{a' b'} \d\q^{\dagger}\g_k\d\q
+ \d_i^{m'}\w^{\ KL}_{m'} \d\q^{\dagger}\g_k \g_{KL}\d\q ] \nn \\
& & + g^{ij}\d_j^{m'}e_{m'}^{\ a'} \d_i^{n'}\w^{\ b'K}_{n'} 
(\d_{a' b'}-i\vare_{a' b'})\d\q^{\dagger} \g_K\d\q .
\eea
Then (\ref{quadtheta}) becomes
\bea
\S_2^{\q} &=& 2 T_S \frac{{Y}}{\pi\rho} \int_{\CC} d^2 \s R V^{-1/3} 
\d\q^{\dagger} \{ \sqrt{g} g^{ij} e_u^{\ k} \del_j y^u 
[ \g_k\del_i -i \d_i^{m'}\w^{\ a' b'}_{m'} \vare_{a' b'} \g_k \nn \\
& & + \d_i^{m'}\w^{\ KL}_{m'} \g_k\g_{KL}]
+ \sqrt{g} g^{ij} \d_j^{m'}e_{m'}^{\ a'} \d_i^{n'}\w^{\ b'K}_{n'} 
(\d_{a' b'}-i\vare_{a' b'}) \g_K \} \d\q .
\eea
As discussed in the previous section, at the level of the quadratic
contributions to the path-integrals, all terms should be evaluated at the 
classical values of the backgound fields. Therefore, the factor 
$2 T_S \frac{{Y}}{\pi\rho} RV^{-1/3}$
can be absorbed by a redefinition of the $\d\q$'s. Next, we use the
relation
\be
\int \cD \d \q \, e^{\int d^2 \s \d \q^{\dagger} \cO \d \q} \propto \det
\cO .
\label{fpathint}
\ee
Note, however, that when going to Euclidean space, we have doubled the number
of fermion degrees of freedom. Therefore, one must actually integrate over
only one half of 
these degrees of freedom. This amounts to taking the square-root of the
determinant on the right-hand side of (\ref{fpathint}). Hence, we conclude
that
\be
\int \cD \d \q \, e^{- \S_2^{\q}} \propto \sqrt{\det \Oslash_3 } ,
\ee
where
\bea
\Oslash_3  &=&  \sqrt{g} g^{ij} \{\g_k e_u^{\ k} \del_j y^u 
[ \del_i -i \d_i^{m'}\w^{\ a' b'}_{m'} \vare_{a' b'} 
+ \d_i^{m'}\w^{\ KL}_{m'} \g_{KL}] \nn \\
& & \ \ \ \ \ \ \ \ \ \ + \g_K \d_j^{m'}e_{m'}^{\ a'} \d_i^{n'}\w^{\ b'K}_{n'} 
(\d_{a' b'}-i\vare_{a' b'}) \} . \label{O3}
\eea 
Note that because of the projections (\ref{thetaugammas}) that 
reduce the number of independent components of $\q$ from 16 to 4, the
operator $\Oslash_3$ must be projected accordingly. We implicitly
assume this.

Collecting the results we have obtained thus far, two-point function
(\ref{pathint5}) can now be written as
\bea
\langle \l_{\bf Y} (y^u_1) \l_{\bf Y} (y^u_2) \rangle \; &\propto &
\frac{\sqrt{\det \Oslash_3 }}{\sqrt{\det \cO_1} } \cdot
\int \cD \l_{\bf Y} \, e^{-\int d^4 y 
\l_{\bf Y} \delslash  \l_{\bf Y}} \l_{\bf Y} (y^u_1) \l_{\bf Y} (y^u_2) \nn \\ 
& & \cdot \int d^4 x\, e^{- \frac{T}{2}{\bf Y}}
\, \l_{\bf Y} (x) \l_{\bf Y} (x) \nn \\ 
& &  \cdot \int \cD \w \,  e^{-S_{0W\!Z\!W}}. \label{pathint7}
\eea
It remains, therefore, to evaluate the $\int \cD \w \, e^{-S_{0W\!Z\!W}}$ 
factor in the path-integral, which we now turn to.

\section{The Wess-Zumino-Witten Determinant:}

In this section, we will discuss the $E_8$ Wess-Zumino-Witten
part of the action, its quadratic expansion and one loop determinant.
Here we follow the exposition in \cite{LOPR} closely. 

Recall from (\ref{boswzw}) that the relevant action is
\bea
S_{0W\!Z\!W} &=& - \frac{1}{8 \pi} \int_{\CC}
 d^2 \s \; \mbox{tr}[\frac{1}{2} \sqrt{g}g^{ij} ( \w_i - A_i )
 \cdot ( {\w}_j - {A}_j ) + i \vare^{ij} {\w}_i {A}_j ]\nn
\\
& & + \frac{1}{24 \pi} \int_{\B} d^3 \hat{\s} i
\hat{\vare}^{\hi\hj\hat{k}}
\Omega_{\hat{k}\hj\hi} (\hat{\w}) , \label{8.1}
\eea
where ${\w}={g}^{-1}d{g}$ is an $E_8 $ Lie algebra
valued
one-form and ${g}$ is given in (\ref{e8}).
In order to discuss the equation of motion and the chirality of this
action, it is convenient to use the complex coordinates $z=\s^0 +i\s^1$,
$\bar{z}=\s^0 -i\s^1$ on $\CC$ and to
define the complex components of ${A}$ by
$A=A_z dz + A_{\bar{z}}d\bar{z}$.
Then action (\ref{8.1}) can be written as
\bea
S_{0W\!Z\!W}&=&-\frac{1}{8\pi}\int_{\CC} d^2 z \;\mbox{tr} \left(
{g}^{-1}
\del_z {g} {g}^{-1} \del_{\bar{z}} {g} -2 {A}_z
{g}^{-1} \del_{\bar{z}} {g} +
{A}_{\bar{z}} {A}_z \right) \nn \\
& & + \frac{1}{24 \pi} \int_{\B} d^3 \hat{\s} i
\hat{\vare}^{\hi\hj\hat{k}}
\Omega_{\hat{k}\hj\hi} (\hat{\w}) . \label{WZWA}
\eea
It is useful to define the two $E_8 $ currents
\be
J_z = (D_z {g}) {g}^{-1}, \ \ \ \ \ \ \ \ \ \ \ \ \ \ \
J_{\bar{z}} = {g}^{-1} D_{\bar{z}} {g} ,
\ee
where $D_z$ and $D_{\bar{z}}$ are the $E_8 $ covariant
derivatives.
In order to perform the path-integral over $\w$, it is necessary to fix
any
residual gauge freedom in the $\w$ fields. Recall from the discussion in
Section 3 that the entire action is invariant under both local gauge and
modified $\k$-transformations, $\d_{\BbL}$ and $\D_{\hat{k}}$
respectively.
It follows from (\ref{gtransf}) and (\ref{gkappa}) that
\be
\d_{\hat{\k}} {g} = {g} i_{\hat{k}} {\BbA} .
\ee
It is not difficult to show that using this transformation, one can choose
a
gauge where
\be
J_z = 0. \label{Jziszero}
\ee
Henceforth, we work in this chiral gauge.
It follows from (\ref{WZWA}) that the ${g}$ equations of motion are
\be
\del_{\bar{z}}J_z=0, \ \ \ \ \ \ \ \ \ \ \ \ \ \ \
D_z J_{\bar{z}}+{F}_{z\bar{z}}=0 , \label{eomfore8}
\ee
where $F_{z\bar{z}}$ is the $E_8 $ field strength. Note
that
this is consistent with the gauge choice (\ref{Jziszero}) and, hence, that
the first equation in (\ref{eomfore8}) is vacuous.
Thus, the on-shell theory we obtain from the gauged Wess-Zumino-Witten
action
is an $E_8$ chiral current algebra
at level one. The level can be read off from the coefficient of the
Chern-Simons
term in (\ref{WZWA}). We would now like to evaluate the Wess-Zumino-Witten
contribution to the path-integral using a saddle-point approximation. To
do
this, we should expand ${g}$ as small fluctuations
\be
{g} = {g}_0 + \d {g}
\ee
around a classical solution ${g}_0$ of (\ref{eomfore8}). However, it
is
clearly rather difficult to carry out the quadratic expansion and evaluate
the
determinant in this formalism. Luckily, there is an equivalent theory
which is
more tractable in this regard, which we now describe.

As discussed in \cite{LOPR}, if the gauge field background is restricted 
to lie within an $SO(16)$ subgroup of $E_8$, 
then the equivalent action is given by the free fermion 
theory coupled to the $SO(16)$ gauge field background. As described in \cite{LOPR}, 
realistic heterotic $M$-theory models can always be chosen to have the gauge 
instanton within the $SO(16)$ subgroup of $E_8$. Here, we consider only 
such restricted backgrounds. We can now write the action
for
$SO(16)$ fermions coupled to background gauge fields. It is
given by \cite{VDP}--\cite{ST}
\be
S_{\psi} = \int_{\CC} d^2 \s  \bar{\psi}^a \Dslash_A^{ab} \psi^b
\ee
where $\psi^a$ denotes the set of $SO(16)$ fermions with
$a,=1,\ldots,16$ and
\be
\Dslash_A^{ab} = \sqrt{g} \t^i (D_i \d^{ab} - A_i^{ab}) 
\label{DA}
\ee
is the covariant derivative on $\CC$ with $A_i^{ab}$ the set of
$SO(16)$ background gauge fields. The matrices $\t^i$ are the Dirac
matrices in two-dimensions. It follows from the above discussion
that we can write
\be
\int \cD \w \, e^{-S_{0W\!Z\!W}} \, \propto \int \cD \psi^a 
\, e^{-S_{\psi}} , \label{wzwpathint}
\ee
where the gauge fixing of variable $\w$ described by (\ref{Jziszero}) is
inherent in the $\psi^a$ formalism, as we will discuss below.
The equations of motion are given by
\be
\Dslash_A^{ab} \psi^b = 0  . \label{eomforpsi}
\ee
We now expand
\be
\psi^a = \psi_0^a + \d \psi^a 
\ee
around a solution $\psi_0^a$ of (\ref{eomforpsi}) and
consider terms in $S_{\psi}$ up to quadratic order in the fluctuations
$\d\psi^a$. We find that
\be
S_{\psi} = S_{0\psi} + S_{2\psi},
\ee
where
\be
S_{0\psi} = \int_{\CC} d^2 \s  \bar{\psi}_0^a \Dslash_A^{ab} \psi_0^b
\ee
and
\be
S_{2\psi} = \int_{\CC} d^2 \s  \d\bar{\psi}^a \Dslash_A^{ab} \d\psi^b
 . \label{S2psi}
\ee
The terms linear in $\d\psi$ vanish by the equations of motion. It follows
immediately from (\ref{eomforpsi}) that $S_{0\psi} = 0$. Then, using
(\ref{fpathint}), one finds from (\ref{S2psi}) that
\be
\int \cD \d \psi^a  \, e^{-S_{\psi}} \, \propto 
\sqrt{\det \Dslash_A}. \label{gaugpathint}
\ee
Note, again, that by going to Euclidean space we have doubled the number of
fermionic degrees of freedom. Therefore, one must actually integrate over only
one half of these degrees of freedom. This requires the square-root of the
determinants to appear in (\ref{gaugpathint}).
It is important to discuss how the chiral gauge fixing condition
(\ref{Jziszero}) is manifested in the $\psi^a$ formalism. Condition 
(\ref{Jziszero}) imposes the constraint that ${g}$ couples only to the
${A}_z$
component of the gauge fields and not to ${A}_{\bar{z}}$. It follows that in
evaluating $\det \Dslash_A $, we should keep only the
${A}_z$ components of the gauge
fields. That is, we should consider the Dirac determinants of
$SO(16)$ holomorphic vector bundles on the Riemann surface $\CC$.
Gauge fixing condition (\ref{Jziszero}) also imposes a constraint on the
definition of determinant $\det \Dslash_A $ as follows.
Recall that on the Euclidean space
$\CC$, each spinor $\psi$ is a complex two-component Weyl spinor
\be
\psi = \left( \ba{c} \psi_+ \\ \psi_- \ea \right) .
\ee
Rescaling this basis to
\be
\left( \ba{c} \psi_+ \\ \psi_- \ea \right) =
\left( \ba{c} (g_{z\bar{z}})^{-1/4} \tilde{\psi}_+ \\
(g_{z\bar{z}})^{1/4} \tilde{\psi}_- \ea \right)
\ee
and using the standard representation for $\t^0,\t^1$ then, locally, one
can
write
\be
\Dslash_A = \left( \ba{cc} 0 & D_{-A} \\ D_{+A} & 0 \ea \right) ,
\label{DAmatrix}
\ee
where
\be
D_{-A} = (g_{z\bar{z}})^{3/4} \left( (g_{z\bar{z}})^{-1/2}
\frac{\del}{\del z}
(g_{z\bar{z}})^{1/2} - A_z \right) , \ \ \ \ \ \ \ \ \ \ \ \ \ \ \
D_{+A} = (g_{z\bar{z}})^{1/4} \frac{\del}{\del \bar{z}} . \label{D-AD+A}
\ee
Since the operator $\Dslash_A$ must be Hermitean, it follows that
$D_{+A}=D_{-A}^{\dagger}$. Now, in addition to disallowing any coupling to
$A_{\bar{z}}$, gauge condition (\ref{Jziszero}) imposes the constraint that
\be
\psi_+^a = 0 \label{psi+iszero}
\ee
for all $a=1,\ldots,16$. Then, using the fact that
\be
\det \Dslash_A = \sqrt{\det (\Dslash_A)^2}
\ee
and gauge condition (\ref{psi+iszero}), we see that the proper definition
of
the determinant is
\be
\det \Dslash_A = \sqrt{\det D_{-A}^{\dagger} D_{-A}} . \label{detDA}
\ee
In this paper, it is not necessary to determine the exact value of
$\det \Dslash_A$.
We need only compute whether it vanishes or is non-zero, and the
conditions
under which these two possibilities occur. To do this, we must examine the
global properties of the holomorphic vector bundle. As we did throughout the 
paper, we will restrict
\be
\CC = \BbC \BbP^1 = S^2 .
\ee
With this restriction, the condition for the vanishing of $\det \Dslash_A$
can
be given explicitly, as we now show.

It follows from (\ref{psi+iszero}) that the chiral fermions realizing the
$SO(16)$ current algebra are elements of the negative chiral spinor line
bundle $S_-$ of the sphere. Note from (\ref{DAmatrix}) that $D_{-A}$
is the part of the Dirac operator which acts on $S_-$. With respect to a
non-trivial $SO(16)$ holomorphic vector bundle background $A$, the complete operator we
should consider is
\be
D_{-A}: S_- \otimes A \to S_+ \otimes A  , \label{D-map}
\ee
where $S_+$ denotes the positive chiral spinor bundle on the sphere. This
is the global description of the local $D_{-A}$ operator defined in
(\ref{DAmatrix}) and (\ref{D-AD+A}).
In order to have nonzero determinant $\det \Dslash_A$, it is necessary and
sufficient
that $D_{-A}$ should not have any zero-modes. This follows from the index
theorem which, for $SO(16)$, implies that
\be
\mbox{coker} D_{-A}^{\dagger} = \mbox{ker} D_{-A} .
\ee
As was shown in \cite{LOPR}, $D_{-A}$ does not have any zero-modes if and only if 
the restriction of the $SO(16)$ holomorphic vector bundle $A$ to $\CC$ is trivial. 
Therefore, in order to have a non-zero superpotential 
for the five-brane, we must have a special type of the gauge bundle on the 
Calabi-Yau threefold. Bundles of this type are straightforward to construct.
We will present a number of phenomenologically relevant examples in a 
forthcoming paper \cite{LOPR2}.

\section{Final Expression for the Superpotential:}

We are now, finally, in a position to extract the final form of the
non-perturbative superpotential from the fermion two-point function.
Combining the results of the previous section with expression
(\ref{pathint7}),
we find that
\bea
\langle \l_{\bf Y} (y^u_1) \l_{\bf Y} (y^u_2) \rangle \; &\propto &
\frac{\sqrt{\det \Oslash_3 }}{\sqrt{\det \cO_1} } \cdot
\sqrt{\det \Dslash_A}  \nn \\
 & & \cdot \int \cD \l_{\bf Y} \, e^{-\int d^4 y 
\l_{{\bf Y}} \delslash  \l_{{\bf Y}}} \l_{\bf Y} (y^u_1) \l_{\bf Y} (y^u_2)
\nn \\
& & \cdot \int d^4 x\, e^{- \frac{T}{2}{\bf Y}(x)}
\, \l_{\bf Y} (x) \l_{\bf Y} (x) .
\eea
Comparing this with the purely holomorphic part of the quadratic fermion
term
in the four-dimensional effective Lagrangian (\ref{4Daction})
\be
(\del_{{\bf Y}} \del_{{\bf Y}} W) \l_{\bf Y} \l_{\bf Y} ,
\ee
we obtain
\be
W \propto \frac{\sqrt{\det \Oslash_3 }}{\sqrt{\det \cO_1} 
}
\cdot \sqrt{\det \Dslash_A}  \, \cdot
e^{-\frac{T}{2}
{\bf Y}}.
\label{final}
\ee
In this expression, the dimensionless field ${\bf Y}$ is defined by
\be
{\bf Y} = \frac{Y}{\pi\rho} \mbox{Re} \cT + i (a + \frac{Y}{\pi\rho}
\mbox{Im}\cT ),
\ee
where $Y$ and $a$ are the translational and axionic moduli of the five-brane
respectively and $\t$ is the complex $(1,1)$-modulus associated with curve
$\CC$. $T$ is a dimensionless parameter given by

\be
T = T_M \pi \rho \ v_{\CC},
\ee
with $T_M$ the membrane tension and $\pi \rho$ the $S^1/\Z_2$ interval
length.
The operators $\cO_1$ and $\Oslash_3 $ are presented in (\ref{Os})
and
(\ref{O3}), respectively. The operator $\Dslash_A$ and its determinant
$\det \Dslash_A$
are defined in (\ref{DA}), (\ref{DAmatrix}), (\ref{D-AD+A}) and
(\ref{detDA}).
This determinant and, hence, the superpotential $W$ will be non-vanishing
if
and only if the pullback of the associated $SO(16)$ 
holomorphic vector bundle $A$ to the
curve
$\CC$ is trivial. 
 All the determinants
contributing to $W$ are non-negative real numbers. We emphasize that $W$
given
in (\ref{final}) is the contribution of open supermembranes wrapped once
around $\CC \times I$, where $\CC = S^2$ is a sphere isolated in
the
Calabi-Yau threefold $CY_3$ and $I \subset S^1/\Z_2$. The generalization to supermembranes wrapped
once
around $I$ but $n$-times around $\CC$ is straightforward. One
simply
replaces the exponential term in (\ref{final}) by
\be
e^{- \frac{nT}{2} {\bf Y}} .
\ee
Further generalizations and discussions of the complete open supermembrane
contributions to the non-perturbative superpotential in heterotic $M$-theory
will be presented elsewhere \cite{LOPR2}.

\appendix

\section{Notation and Conventions:}

We use a notation such that symbols and indices without hats represent 
fields in the ten-dimensional fixed hyperplanes of \HW theory 
(as well as the two-dimensional heterotic string theory), 
while hatted indices relate to quantities of eleven-dimensional bulk space
(and the three-dimensional open membrane theory). In addition, underlined
symbols and indices refer to the five-brane worldvolume.

\subsection*{Bosons:}
    
For example,
\be
X^M, \ \ \ M=0,1,\ldots,9, \ \ \ \ \mbox{and} \ \ \ \ \hat{X}^{\hat{M}},
\ \ \ \hat{M}=\hat{0},\hat{1},\ldots,\hat{9},\hat{11}, \label{app1}
\ee
are, respectively, the coordinates of ten- and eleven-dimensional spacetimes. 
We do not change notation when switching from Minkowskian signature to 
Euclidean signature.

Eleven-dimensional space is, by assumption, given by
\be
M_{11} = R_4 \times CY_3 \times S^1/\Z_2 ,
\ee
while the ten-dimensional space obtained by compactifying it on $S^1/\Z_2$ is, 
clearly,
\be
M_{10} = R_4 \times CY_3 .
\ee
The membrane worldvolume $\Si$ is decomposed as
\be
\Si = \CC \times I ,
\ee
where the holomorphic curve $\CC$ lies within $CY_3$ and $I \subset S^1/\Z_2$.

The two-dimensional heterotic string theory is represented by fields with
worldsheet coordinates $\s^i$, with $i=0,1$. Bosonic indices of ten-dimensional
spacetime are split into indices parallel to the worldsheet
($m' = 0,1$) and indices perpendicular to it ($m''=2,\ldots,9$).
The space normal to the worldsheet is an eight-dimensional space. Since
it is assumed that the worldsheet is wrapped on a curve $\CC$ contained in 
the Calabi-Yau 
threefold $CY_3$, these eight directions $y^{m''}$ can be split in two sets 
of four. The first set parametrizes the subset $CY_{\perp} \subset CY_3$ which 
is normal to curve $\CC$. The coordinates are denoted $y^U$, $U=6,7,8,9$.
The second set consists of the coordinates $y^u$, $u=2,3,4,5$ of $R_4$. 

The five-brane worldvolume $\ul{M}_6$ is embedded in $M_{10}$ as
\be
M_{10} = CY_{\perp} \times \ul{M}_6 ,
\ee
where 
\be
\ul{M}_6 = R_4 \times \CC .
\ee
Indices in the five-brane super-worldvolume are given by 
$\ubR = (\ul{r},\ul{\m})$, with $\ul{r}=0,1,\dots,5$ and 
$\ul{\m}=1,\ldots,16$. Note that one has a $(2,0)$-supersymmetry
on the five-brane worldvolume.

Coordinates of $CY_3$ are denoted by
\be
\breve{y}^{\breve{U}} = (X^{\ul{r}'},y^U) , \ \ \ \ \ \ \ \mbox{with} \ \ \ 
\breve{U}=0,1,6,7,8,9,\ \ \ \ \ul{r}'=0,1,
\ee
or, using the complex structure notation, 
\be
\breve{y}^m, \ \ \ \breve{y}^{\bar{m}}, \ \ \ \ \ \ \ \ \ \ \ \
m= 1,2,3, \ \ \ \ \bar{m}=\bar{1},\bar{2},\bar{3}.    \label{app6}
\ee

The bosonic indices in (\ref{app1})-(\ref{app6}) are coordinate 
(or ``curved'') indices. The corresponding
tangent space (or ``flat'') indices are given in the following table,

\begin{center}
\begin{tabular}{||c|c|c|c|c|c|c||} \hline
$M_{10}$ & $M_{11}$ & $\ul{M}_6$ & $\CC$ & $M_{\perp}$ & $R^4$ & $CY_{\perp}$ \\ \hline 
\hline
$M,N$ & $\hat{M},\hat{N}$ & $\ul{r},\ul{s}$ & $\ul{r}',\ul{s}'$ & $m'',n''$ & $u,v$ & $U,V$ \\ \hline
$A,B$ & $\hat{A},\hat{B}$ & $\ul{a},\ul{b}$ & $\ul{a}',\ul{b}'$ & $a'',b''$ & $k,l$ & $K,L$ \\ \hline
\end{tabular}
\end{center}

\noindent where $M_{\perp}$ is the subspace of $M_{10}$ perpendicular to $\CC$.

\subsection*{Spinors:}
    
In ten-dimensional spacetime with Euclidean signature, the $32\times 32$ 
Dirac matrices $\G_A$ satisfy
\be
\{ \G_A, \G_B \} = 2 \eta_{A B}
\ee
or, with curved indices, (since $\G_A = e_A^{\ M} \G_M$)
\be
\{ \G_M, \G_N \} = 2 g_{M N}.
\ee
One defines ten-dimensional chirality projection operators 
$\frac{1}{2} (1\pm \G_{11})$,
where
\be
\G_{11} = - i \G_0 \G_1 \cdots \G_9.
\ee
A useful representation for $\G_A$ is given by the two-eight split
\be
\G_{A} = ( \t_{a'} \otimes \tilde{\g} , 1 \otimes \g_{a''} ) ,
\ee
where the two-dimensional Dirac matrices $\t_0,\t_1$ and their product defined
by $\tilde{\t}= -i\t_0 \t_1$ are explicitly given by
\be
\t_0 = \left( \ba{cc} 0 & 1 \\ 1 & 0 \ea \right) , \ \ \ 
\t_1 = \left( \ba{cc} 0 & -i \\ i & 0 \ea \right) , \ \ \ 
\tilde{\t} = \left( \ba{cc} 1 & 0 \\ 0 & -1 \ea \right). 
\ee
These ten-dimensional Dirac matrices are more explicitly written as
\be
\ba{cc}
\G_0 = \left( \ba{cc} 0 & \tilde{\g} \\ \tilde{\g} & 0 \ea \right) &
\G_1 = \left( \ba{cc} 0 & -i\tilde{\g} \\ i\tilde{\g} & 0 \ea \right) \\
\G_{a''} = \left( \ba{cc} \g_{a''} & 0 \\ 0 & \g_{a''} \ea \right) &
\G_{11} = \left( \ba{cc} \tilde{\g} & 0 \\ 0 & -\tilde{\g} \ea \right) \ea
\ee
where $\g_{a''}$ are $16\times 16$ Dirac matrices, and the product
\be
\tilde{\g} = \g_2 \g_3 \cdots \g_9
\ee
is used in the definition of eight-dimensional chirality projection operators 
$\frac{1}{2} (1\pm \tilde{\g})$.
Note that $\G_{11}^2 = 1$, $ \tilde{\g}^2 = 1 $, and $ \tilde{\t}^2 = 1$. 
In eleven-dimensions, the $32 \times 32$ Dirac matrices are given by
\be
\hat{\G}_{\hat{A}} = \G_{\hat{A}}, \ \ (\hat{A}=\hat{0},\hat{1},\ldots,\hat{9}),
\ \ \ \ \ \mbox{and} \ \ \ \ \ \hat{\G}_{\hat{11}} = \G_{11}.
\ee


\begin{thebibliography}{99}

\bibitem{HorWit1}P. Ho\v{r}ava and E. Witten, {\sl ``Heterotic and Type I
String
Dynamics from Eleven Dimensions''}, Nucl. Phys. {\bf B460} (1996) 506, 
hep-th/9603142.

\bibitem{HorWit2}P. Ho\v{r}ava and E. Witten, {\sl ``Eleven-Dimensional
Supergravity on a Manifold with Boundary''}, Nucl. Phys. {\bf B475} (1996)
94,
hep-th/9510209.

\bibitem{W97}E. Witten, {\sl ``Solutions of Four-Dimensional Field Theories
via $M$-Theory''}, Nucl. Phys. {\bf B507} (1997) 3, hep-th/9703166.

\bibitem{c10}T. Banks and M. Dine, Nucl. Phys. {\bf B479} (1996) 173.

\bibitem{BAD}A. Lukas, B.A. Ovrut and D. Waldram, {\sl ``On the Four-Dimensional
Effective Action of Strongly Coupled Heterotic String Theory''}, Nucl. Phys.
{\bf B532} (1998) 43, hep-th/9710208.

\bibitem{B34}A. Lukas, B.A. Ovrut, K. Stelle and D. Waldram, {\sl ``The
Universe
as a Domain Wall''}, Phys. Rev. {\bf D59} (1999) 086001, hep-th/9803235.

\bibitem{B33}A. Lukas, B.A. Ovrut and D. Waldram, {\sl ``Cosmological
Solutions
of \HW Theory''}, Phys. Rev. {\bf D60} (1999) 086001, hep-th/9806022.

\bibitem{B32}A. Lukas, B.A. Ovrut, K. Stelle and D. Waldram, {\sl
``Heterotic
$M$-theory in Five Dimensions''}, Nucl. Phys. {\bf B552} (1999) 246,
hep-th/9806051.

\bibitem{B31}A. Lukas, B.A. Ovrut and D. Waldram, {\sl ``Non-Standard
Embedding
and Five-Branes in Heterotic $M$-theory''}, Phys. Rev. {\bf D59} (1999)
106005, 
hep-th/9808101.

\bibitem{B27}R. Donagi, T. Pantev, B.A. Ovrut and D. Waldram, {\sl
``Holomorphic
Vector Bundles and Nonperturbative Vacua in $M$-theory''}, JHEP {\bf 9906}
(1999) 
34, hep-th/9901009.

\bibitem{B15}R. Donagi, B.A. Ovrut, T. Pantev and D. Waldram, {\sl
``Standard
Models from Heterotic $M$-theory''}, hep-th/9912208.

\bibitem{B5}R. Donagi, B.A. Ovrut, T. Pantev and D. Waldram, {\sl
``Standard
Model Bundles on Nonsimply Connected Calabi-Yau Threefolds''},
hep-th/0008008.

\bibitem{B4}R. Donagi, B.A. Ovrut, T. Pantev and D. Waldram, {\sl
``Standard
Model Bundles''}, math-ag/0008010.

\bibitem{B3}R. Donagi, B.A. Ovrut, T. Pantev and D. Waldram, {\sl
``Spectral
Involutions on Rational Elliptic Surfaces''}, math-ag/0008011.

\bibitem{c32}H.P. Nilles, M. Olechowski and M. Yamaguchi, {\sl
``Supersymmetry
Breakdown at a Hidden Wall''}, hep-th/9801030.

\bibitem{c39}E. Sharpe, {\sl ``Boundary Superpotentials''},
hep-th/9611196.

\bibitem{c40}E.A. Mirabelli and M.E. Peskin, {\sl ``Transmission of
Supersymmetry
Breaking from a Four-dimensional Boundary''}, hep-th/9712214.

\bibitem{RS}L. Randall and R. Sundrum, Phys. Rev. Lett. {\bf 83} (1999)
3370, 
hep-th/9905221.

\bibitem{RS'}L. Randall and R. Sundrum, Phys. Rev. Lett. {\bf 83} (1999)
4690, 
hep-th/9906064.

\bibitem{c41}A. Lukas and B.A. Ovrut, {\sl ``U-Duality Invariant $M$-theory
Cosmology''}, hep-th/9709030.

\bibitem{c42}K. Benakli, {\sl ``Cosmological Solutions in $M$-theory on
$S^1/\Z_2$''},
hep-th/9804096.

\bibitem{c43}K. Benakli, {\sl ``Scales and Cosmological Application in
$M$-theory''},
hep-th/9805181.

\bibitem{c44}A. Lukas, B.A. Ovrut and D. Waldram, {\sl ``Cosmological
Solutions
of \HW Theory''}, Phys. Rev. {\bf D60} (1999) 086001, hep-th/9806022.

\bibitem{c45}J. Ellis, Z. Lalak, S. Pokorski and W. Pokorski, 
{\sl ``Five-Dimensional Aspects of $M$-theory and Supersymmetry Breaking''}, 
hep-th/9805377.

\bibitem{LOPR}E. Lima, B. Ovut, J. Park and R. Reinbacher, {\sl 
``Non-Perturbative Superpotential from Membrane Instantons in
Heterotic $M$-theory''}, hep-th/0101049.

\bibitem{BeckerBS}K. Becker, M. Becker and A. Strominger, 
{\sl ``Fivebranes, Membranes and Non Perturbative String Theory''}, Nucl.
Phys. 
{\bf B456} (1995) 130, hep-th/9507158.

\bibitem{HarvMoor}J. Harvey and G. Moore, {\sl ``Superpotentials and
Membrane
Instantons''}, hep-th/9907026.

\bibitem{Derend}J.-P. Derendinger and R. Sauser, {\sl ``A Five-brane Modulus
in the Effective N=1 Supergravity of $M$-theory''}, hep-th/0009054.

\bibitem{MPS}G. Moore, G. Peradze and N. Saulina, {\sl ``Instabilities in 
Heterotic $M$-theory Induced by Open Membrane Instantons''}, hep-th/0012104.

\bibitem{CJS}E. Cremmer, B. Julia and J. Scherk, Phys. Lett. {\bf B76} (1978)
409.

\bibitem{DuffStelle}M.J. Duff and K.S. Stelle, Phys. Lett. {\bf B253}
(1991) 
113.

\bibitem{BST}E. Bergshoeff, E. Sezgin and P.K. Townsend, Phys. Lett. {\bf
B189}
(1987) 75.

\bibitem{BLNPST}I. Bandos, K. Lechner, A. Nurmagambetov, P. Pasti,
D. Sorokin and M. Tonin, Phys. Rev. Lett. {\bf 78} (1997) 4332.

\bibitem{cremmferr}E. Cremmer and S. Ferrara, Phys. Lett. {\bf B91} (1980)
61.

\bibitem{Ceder}M. Cederwall, {\sl ``Boundaries of 11-Dimensional
Membranes''}, 
hep-th/9704161.

\bibitem{LalLukOvr}Z. Lalak, A. Lukas and B.A. Ovrut, {\sl ``Soliton
Solutions
of $M$-theory on an Orbifold''}, hep-th/9709214.

\bibitem{Cand}P. Candelas, {\sl Lectures on Complex Manifolds}, in Trieste
1987 Proceedings.

\bibitem{Wnab}E. Witten, Comm. Math. Phys. {\bf 92} (1984) 455.

\bibitem{chusezgin}C.S. Chu and E. Sezgin, 
{\sl ``M-Fivebrane from the Open Supermembrane''}, JHEP {\bf 9712} (1997) 1.

\bibitem{burtlec}B. Ovrut, {\sl ``N=1 Supersymmetric Vacua in
Heterotic $M$-theory''}, hep-th/9905115.

\bibitem{dewitetal}B. de Wit, K. Peeters and J. Plefka, {\sl ``Superspace 
Geometry for Supermembrane Backgrounds''}, hep-th/9803209.

\bibitem{Berg}E. Bergshoeff, M. de Roo, B. de Wit and P. van
Nieuwenhuizen,
Nucl. Phys. {\bf B195} (1982) 97.

\bibitem{chernlec}S.S. Chern, {\sl ``Minimal Submanifolds in a Riemannian
Manifold''}, University of Kansas, Dept. of Mathematics, 1968.

\bibitem{GSW1}M. Green, J. Schwartz and E. Witten, {\sl Superstring Theory, 
Vol.1}, Cambridge University Press, 1987.

\bibitem{AGetal}L. Alvarez-Goum\'{e}, G. Moore and C. Vafa, Comm. Math.
Phys. 
{\bf 106} (1986) 81.

\bibitem{Freed}D. Freed, {\sl ``On Determinant Line Bundles''}, in
Mathematical
Aspects of String Theory, edited by S. Yau (World Scientific, 1987).

\bibitem{VDP}D.Vecchia, B. Durhuus and J.L. Petersen, Phys. Lett. {\bf
B144} (1984) 245.

\bibitem{RS1}A.N. Redlich and H. Schnitzer, Nucl. Phys. {\bf B183} (1987)
183.

\bibitem{RS2}A.N. Redlich and H. Schnitzer, Nucl. Phys. {\bf B167} (1986)
315.

\bibitem{RST}A.N. Redlich, H. Schnitzer and K. Tsokos, Nucl. Phys. {\bf
B289} 
(1989) 397.

\bibitem{ST}H. Schnitzer and K. Tsokos, Nucl. Phys. {\bf B291} (1989) 429.

\bibitem{LOPR2}E. Lima, B.A. Ovrut, J. Park and R. Reinbacher, In
preparation.


\end{thebibliography}
\end{document}